\documentclass{aa}  

\usepackage{graphicx}
\usepackage{lscape}
\usepackage{txfonts}
\begin{document}

   \title{Formation of supermassive stars in the first stellar clusters: Dependence on the gas temperature}

   \author{P.A. Solar
          \inst{1, 2}
          \and
          B. Reinoso\inst{3,5}
\and
D.R.G. Schleicher\inst{2}
          \and
          R.S. Klessen\inst{3,4}
          \and 
          Robi Banerjee\inst{1}
          }

   \institute{$^1$Hamburg Observatory, Hamburg University, Gojenbergsweg 112, 21029 Hamburg, Germany              \email{paulo.solar.vera@uni-hamburg.de}    \\
             $^2$Departamento de Astronom\'ia, Facultad Ciencias F\'isicas y Matem\'aticas, Universidad de Concepci\'on, Av. Esteban Iturra s/n, Concepci\'on, Chile\\
$^3$Universit\"at Heidelberg, Zentrum f\"ur Astronomie, Institut f\"ur Theoretische Astrophysik, Albert-Ueberle-Str. 2, 69120 Heidelberg, Germany\\
$^4$Universit\"{a}t Heidelberg, Interdisziplin\"{a}res Zentrum f\"{u}r Wissenschaftliches Rechnen, Im Neuenheimer Feld 205, 69120 Heidelberg, Germany\\
$^5$Department of Physics, Gustaf H\"allstr\"omin katu 2, FI-00014, University of Helsinki, Finland
             }

   \date{Received May 25, 2024; accepted aMarch 16, 1997}

  \abstract 
   {The origin of supermassive black holes is an open question that has been explored considering gas- and collision-based formation channels to explain the high number of quasars observed in the early Universe. According to numerical simulations, supermassive stars can be formed in atomic cooling halos when protostars reach accretion rates greater than $\sim 10^{-2}~\mathrm{M_{\odot}~yr^{-1}}$ and fragmentation is inhibited on parsec scales. It remains uncertain, however, whether fragmentation on smaller scales leads to the formation of a star cluster instead of a supermassive star in the presence of possible cooling mechanisms.}
  % aims heading (mandatory)
   {We explored the formation of a central massive object through collisions and the accretion of Population III stars in a primordial gas cloud in a gravitationally unstable system by varying the gas temperature and thus the degree of gravitational instability. We explored the impact of disk fragmentation and compared our results with theoretical accretion rates.}
  % methods heading (mandatory)
   {We evolved a small Population III star cluster embedded in a primordial gas cloud on subparsec scales considering a gravitationally unstable initial configuration with different gas temperatures.
   We performed multiphysics simulations in the AMUSE framework with a hydrodynamical gas treatment through Smoothed-particle hydrodynamics and $N$-body dynamics for the protostars represented through sink particles. To do this, we incorporated physically motivated accretion recipes. We also included a realistic mass-radius relation and solved the collisions with the sticky-sphere approximation.}
  % results heading (mandatory)
   {Our results show that central massive objects with masses $\sim 10^4~\mathrm{M_{\odot}}$ can be formed by accretion and collisions at different temperatures and that the most massive object can reach efficiencies of $\sim 0.61$ for atomic cooling conditions and $\sim 0.95$ for more unstable conditions. We observe a quasi-disk formation for warmer temperatures and a higher contribution through collisions to the mass of a central massive object. Our results show that the embedded cluster is in a supercompetitive accretion regime in which it obtains mass by accretion that is regulated by self-gravity at the beginning and via Bondi–Hoyle accretion at later times in simulations with higher temperatures.}
  % conclusions heading (optional), leave it empty if necessary 
   {Our results suggest that in more unstable conditions with lower gas temperatures, a more massive supermassive black hole seed can form. This corresponds to a higher efficiency in the formation of the central object.}

   \keywords{methods: numerical – stars: formation – quasars: supermassive black holes – stars: Population III – early Universe.
               }

   \maketitle
%
%-------------------------------------------------------------------

\section{Introduction}

More than 200 supermassive black holes (SMBHs) have been detected at $z>5.7$ \citep{Fan2006,Mortlock2011,Wu15Nature,Banados2018,Reed2019,Onoue2019,Banados2021,Wang21}, with masses higher than $10^9~\rm M_\odot$. This also includes SMBHs as massive as $10^{10}~\rm M_\odot$ for SDSS J010013.02+280225.8 \citep{Wu15Nature} or from the recent new observations of the \textit{James Webb Space Telescope} (JWST) in collaboration with the \textit{Chandra X-ray Observatory}, which is the currently highest-redshift quasar UHZ1 at $z=10.3$ with a mass of about $\mathrm{M_{bh} \sim 10^7 - 10^8~M_{\odot}}$, as reported by \citet{2024uhz1}. The possible origin of these objects was discussed early by \citet{Rees1984}, who also discussed a variety of possible formation pathways, including the possibility of a direct collapse, in which a massive gas cloud collapses directly into one or a small number of objects \citep[e.g.,][]{Bromm2003,Koushiappas2004,Wise2008,Begelman2009,Schleicher2010,Latif2013}. A viable alternative for the formation of SMBHs is the formation of very massive objects through collisions in dense stellar clusters \citep{Omukai2008,Devecchi2009,Katz2015,Sakurai2017,Sakurai2019,Reinoso18,Reinoso2020,Vergara21, Escala2021, Vergara2022,2023vergara} or black hole mergers in dense black hole clusters \citep{Davies11,Lupi14}.

Another frequently discussed possibility is that the first SMBHs may originate from the first stars in the Universe \citep{Abel2002,Heger2002,Heger2003,Klessen19} through their top-heavy initial mass function \citep{Clark2011, Greif2011,Susa2014,Stacy2016,Fraser2017,Rafeel18,Sharda2020}. In comparison to the other scenarios these seeds would be relatively light, however. Another challenge is that these stars very effectively removing gas from the low-mass halos through radiative and supernova feedback. This impedes subsequent growth \citep{Johnson2007,Smith2018}.

More recent investigations have also suggested that a realistic pathway for the formation of a very massive object may not only consist of the idealized scenarios originally proposed by \citet{Rees1984}, but that several variations are also possible, including hybrid scenarios that combine gas and stellar dynamical effects. For example, while cosmological simulations found that fragmentation occurs in gas dynamical studies, the resulting clumps may also merge again and do not necessarily inhibit the formation of a very massive object \citep[see, e.g.,][]{Grete2019, Suazo2019,2023regan,2018regan2,2018regan,2024prole}. It is well known that gravitational torques in self-gravitating disks may produce exactly this type of behavior \citep{LatifSchleicher2015}. Similarly, the very high accretion rates that were often found in gas dynamical simulations are expected to enhance the radii of the protostars, which in turn enhances their probability for collisions \citep[e.g.,][]{Hosokawa2013, Schleicher2013, Haemmerle2018, Haemmerle2019, Haemmerle2021}.

A more systematic exploration of these hybrid scenarios has been pursued by 
\citet{Boekholt2018}, who presented toy models to study the interplay of accretion and collisions. The authors showed that the latter could lead to the formation of very massive central objects in dense compact clusters. \citet{Seguel2020} included the mass loss during collisions of protostars in these simulations. \citet{Tagawa2020} developed a semianalytic model considering stellar bombardment and accretion, while \citet{Das2021} showed that the collision probability is enhanced in the presence of accretion through the absence of mass and linear momentum conservation.

The question regarding the origin remains very important today. The original formation needs to be taken into account together with subsequent accretion and other processes, such as the growth via merger events \citep[see, e.g., the reviews by][]{Volonteri10,review_woods19}. Semianalytic models that include these processes are very helpful because they provide an assessment of the black hole population that results from different assumptions on the different seeds. This provided some tentative indications that hybrid models that might work at very low but nonzero metallicities may be an important ingredient to consider \citep{2016Habouzit,Sassano2021,Trinca2022}. \citet{Chon2020} presented simulations that incorporated dust cooling and showed that the resulting fragmentation does not have to impede the formation of a massive object. An analytic scenario showing that collisions and accretion can be expected to interact and form massive objects was presented by \citet{Schleicher2022}. \citet{Reinoso2023} presented a suite of numerical simulations that consistently modeled the gas and $N$-body dynamics of the protostars in a compact cluster, demonstrating in detail how these processes interact and lead to the formation of very massive objects.

We follow up on the recent study by \citet{Reinoso2023}, with the objective to study even more unstable systems, that is, systems with a higher ratio of the gas mass to the Jeans mass, because we expect this regime to be particularly relevant and important for the formation of very massive objects. For this purpose, we varied the gas temperature from $8000~\mathrm{K}$, where atomic hydrogen cooling dominates, to $500~\mathrm{K}$, where $\mathrm{H_2}$ cooling and dust thermal emission become efficient and where gravitational
instability might be increased. Our numerical approach is similar to that of \citet{Reinoso2023} and is briefly presented in section~\ref{method} for completeness. Our results are given in section~\ref{results}, and we discuss our main conclusions in section~\ref{discussion}.

\section{Numerical method}\label{method}

To explore how the gas-dynamics and protostellar dynamics affect the formation of a SMBH seed we model a protostellar cluster inside a compact gas cloud in virial equilibrium. The methodology adopted here is based on the Astrophysical MUlti-purpose Software Environment \citep[AMUSE\footnote{https://github.com/amusecode/amuse}, see][]{AMUSE_Portegies09,AMUSE_Portegies13,AMUSE_Pelupessy13,Portegies2018}, a {\small PYTHON} interface designed to couple existing numerical codes. We closely follow the methodology presented by \citet{Reinoso2023}, including mass accretion onto protostars, sink particle creation, a treatment for stellar collisions, and a mass-radius relation for the protostars.

\begin{table*}
\caption{Summary of the initial conditions employed in our simulations.}
\begin{center}
\begin{tabular}{ c c c c c c c c }
\hline\hline
$\mathrm{N_{SPH}}$ & $\mathrm{M_{gas}}$ & $\mathrm{R_{gas}}$ & $\mathrm{M_{1,gas}}$ & T & $\mathrm{N_{star}}$ & $\mathrm{M_{star}}$ & $\mathrm{M_{1,star}}$  \\ 
$~$ & $\mathrm{[M_{\odot}]}$ & $\mathrm{[pc]}$ & $\mathrm{[M_{\odot}]}$ & [K] & $~$& $\mathrm{[M_{\odot}]}$ & $\mathrm{[M_{\odot}]}$ \\ \hline
$2^{18}$ & $3\times10^{4}$ & $0.14$ & $\sim 0.11$ & 500 & 256 & 25.6 & 0.1 \\
$2^{18}$ & $3\times10^{4}$ & $0.14$ & $\sim 0.11$ & 1000 & 256 & 25.6 & 0.1 \\
$2^{18}$ & $3\times10^{4}$ & $0.14$ & $\sim 0.11$ & 3000 & 256 & 25.6 & 0.1 \\
$2^{18}$ & $3\times10^{4}$ & $0.14$ & $\sim 0.11$ & 5000 & 256 & 25.6 & 0.1 \\
$2^{18}$ & $3\times10^{4}$ & $0.14$ & $\sim 0.11$ & 8000 & 256 & 25.6 & 0.1 \\ \hline
\end{tabular}
\tablefoot{We show the number of SPH particles used for each simulation. The input parameters are the initial gas mass $\mathrm{M_{gas}}$, the initial gas radius $\mathrm{R_{gas}}$, the approximate initial gas mass for each SPH particle $\mathrm{M_{1,gas}}$, the initial temperature $\mathrm{T}$, the number of stars $\mathrm{N_{stars}}$, the total stellar mass $\mathrm{M_{star}}$, and the mass of each star $\mathrm{M_{1,star}}$. }
\label{tab:initial_condition}
\end{center}
\end{table*}

The $N$-body code {\small PH4} \citep{McMillan96} and the Smoothed-particle hydrodynamics (SPH) code {\small FI} \citep{Hernquist_Katz1989,Gerritsen1997,Pelupessy2004} are coupled via the {\small BRIDGE} method \citep{Fujii2007}  in {\small AMUSE}. The gravitational acceleration at the position of the $N$-body particles is calculated using the SPH particles and vice versa. To ensure that the coupling does not violate Newton's third law, we employ the code {\small FastKick}  with a constant gravitational smoothing length of $0.5~\mathrm{au}$ corresponding to the smallest smoothing length in the simulations. The particles in the $N$-body code are solved with a smaller smoothing length of $1~\mathrm{R_\odot}$ for a very accurate treatment of the gravitational interactions between the protostars. An external pressure floor is included in the SPH code using the treatment of \citet{Benz1990,Clark2011}.
The external pressure is set equal to the pressure of the cloud at the cut-off radius, corresponding to $1.78~\times 10^{-7}~\mathrm{g~cm^{-1}~s^{-2}}$ for the hottest simulation and $1.13~\times 10^{-8}~\mathrm{g~cm^{-1}~s^{-2}}$ for the coldest one.

We employ a modified equation of state of the form
\begin{equation}
\label{eq:EOS}
    T = T_0 \left[ 1 + \left( \frac{\rho}{\rho_{\rm c}} \right)^{\gamma - 1} \right],
\end{equation}
such that the gas behaves isothermally, with temperatures from $T_0=500 ~\rm K - 8000~\rm K$ at low densities, but becomes adiabatic at densities above $\rho_{\rm c}=10^{15}~\mathrm{cm^{-3}}$, as found in 1D and 3D models including detailed chemical networks with $\gamma=5/3$ \citep{Omukai2008,Becerra2015}.\\

The time integration is based on the Kick-Drift-Kick (KDK) integration with \textit{bridge} including protostellar collisions and sink particle creation using the algorithms described below. We employ a maximum timestep of 5~yr for \textit{bridge} along with an adaptive time-stepping algorithm, where the timestep is reduced by factors of 2 until becoming smaller than the shortest free-fall time among all the SPH particles. The $N$-body particles in our simulations represent protostars. Using the particle sets in {\small AMUSE}, we assign additional properties  such as: \textit{stage} and \textit{luminosity}. The \textit{stage} property indicates the phase of stellar evolution, that is, when the object has already entered the main sequence phase or if it is still in the protostellar phase, allowing to take into account the difference in the protostellar radii, which is highly relevant for collisions. We use a mass-radius relationship according to \citet{Reinoso2023}, based on the works by \citet{Hosokawa09,Hosokawa12}, and \citet{Hosokawa2013}. The $N$-body particles evolve through three different stages: protostars, star and supermassive star (SMS). When a particle is in the \textit{star stage}, it grows as a main sequence star according to the following mass-radius relation:

\begin{equation} 
R_{\rm star} = 0.97\left(\frac{M_{\rm star}}{\mathrm{M_{\odot}}}\right)^{0.57}~\mathrm{R_{\odot}}.
\end{equation}

Particles in a \textit{SMS stage} have an accretion rate greater than the critical accretion rate $\dot{m}_{\rm crit} = 0.04~ \mathrm{M_{\odot
}~yr^{-1}} $ \citep{Hosokawa2013}. The mass-radius relation of a particle in a \textit{protostar stage} is determined by the accretion rate $\dot{m}$, where the swelling or contraction of the protostar will bring it closer to either the \textit{star} or \textit{SMS stage}, depending on the accretion rate. A protostar can contract to a \textit{star stage} if its accretion rate falls below the critical accretion rate for a period of time longer than a factor $100$ of the Kelvin-Helmholtz timescale ($100t_{\rm KH}$). For a detailed description of the mass-radius relation, see Appendix A in \citet{Reinoso2023}.

After each KDK step, we calculate the gas accretion onto the protostars employing the algorithm from \citet{Hubber13}. We consider a spherical volume called the interaction zone with radius $R_{\rm I.Z.}$. Inside this radius the weighted average gas flux onto the central point mass is computed using a cubic spline kernel function. The interaction radius is adjusted iteratively with a maximum of 50 iterations to maintain a constant gas mass of $M_{\rm int, max}=50$~$M_{\rm gas}/{N_{\rm SPH}}$ (the mass corresponding to 50 SPH particles). We further impose that the interaction zone cannot be smaller than the protostars. We employ angular momentum conservation for the accreted gas and include the angular momentum feedback described by \citet
{Hubber13}, including their prescriptions for accretion. We use their NewSink algorithm to create sink particles when the following conditions are fulfilled: i) an SPH particle reaches a density higher than $10^{16}~\mathrm{cm^{-3}}$ \citep{Becerra2015}; ii) it sits in a minimum of the gravitational potential among its neighbors; iii) it does not overlap with existing sinks; iv) it fulfills the density criterion:
\begin{equation}
    \rho_{i} > \rho_{\rm Hill} \equiv \frac{3 X_{\rm Hill} (-\Delta\mathrm{r}_{is'} \cdot \Delta\mathrm{a}_{is'} )}{4\pi G |\Delta\mathrm{r}_{is'}|^2},
\end{equation}
for all existing sinks $s'$ for a given SPH particle $i$. Here, $\Delta\mathbf{r}_{is'}$ and $\Delta\mathbf{a}_{is'}$ are the relative position and acceleration of sink candidate $i$ with respect to an existing sink $s'$, with  $X_{\rm Hill}=4$. The last criterion is based on the Hill sphere and ensures that an SPH particle turns into a sink particle in the vicinity of another sink only if the density peak dominates the local gravitational potential.

A collision between two particles occurs when their radii overlap during the $N$-body integration, that is,
\begin{equation*}
    d \leq R_1 + R_2,
\end{equation*}
where $d$ is the separation between the particles and $R_1$ and $R_2$ are their radii. We then replace the overlapping particles with a new particle placed at the center of mass where the new velocity is calculated following momentum conservation. The new mass of the particle is the sum of the masses of the colliding particles, that is,
\begin{equation*}
M_{\rm new} = M_1 + M_2.
\end{equation*}

\section{Initial conditions}\label{initial conditions}

The gas is modeled  through SPH particles, with a total of $2^{18}$ SPH particles per simulation. In all simulations the mass of the gas cloud is $M_{\rm gas} = 3\times 10^4 ~ \mathrm{M_{\odot}}$, with a virial radius of $0.14~ \mathrm{pc}$ . The protostars were simulated by sink particles that interact only through gravity with themselves and with the gas. The initial number of sink particles is 256, each with a mass of $0.1~ \mathrm{M_{\odot}}$, which are compatible with protostellar masses that emerged from atomic cooling halos \citep{Becerra2015}. Both the protostars and the gas particles follow a plummer distribution \citep{plummer1911}, with a half-mass radius of $0.1~ \mathrm{pc}$ and a cut-off radius of 5 Plummer radii for each model. The initial velocity of the gas particles is zero and for the protostars the velocities follow a virial equilibrium condition. To be consistent with the results found by \citet{Latif2013}, we use the same recipe as \citet{Reinoso2023} to inject a spectrum of noncompressive Kolmogorov turbulence with Mach number $\mathcal{M} = 1$. We simulate isothermal gas clouds with different initial temperatures and we assume a primordial composition of 75\% atomic hydrogen and 25\% helium. In Table~\ref{tab:initial_condition} we summarize the initial conditions for all simulations performed in this study.

\section{Results}\label{results}
In this section we present the main results of our simulations, adopting the initial conditions  from Table~\ref{tab:initial_condition}, where we vary the initial temperature to explore its impact on the formation of the most massive object (MMO) in the cluster. The simulations are stopped at approximately $\sim 30000~\mathrm{yr}$,  because at this point most of the gas has already been accreted and a supermassive object at the center of the cluster is clearly established. 

\begin{figure*}
    \centering
	\includegraphics[width=17cm]{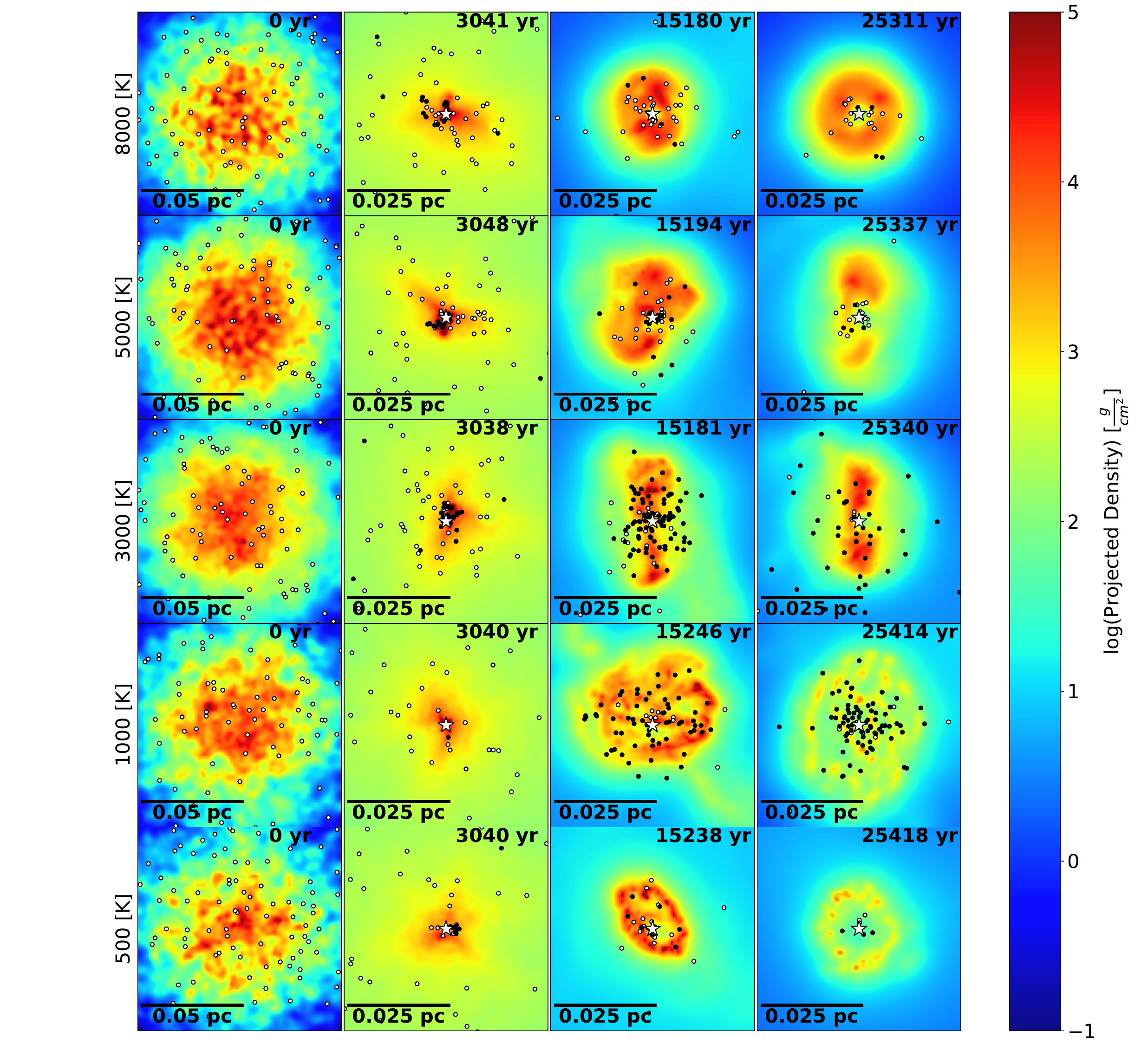}
    \caption{Density projections of the face-on view of a primordial gas cloud with an embedded protostar cluster for simulations with an initial temperature of $500~\mathrm{K}-8000~\mathrm{K}$ at different times. The white dots are the protostars, black dots are the formed stars and the white star is the MMO. The clusters evolve in time from left to right and from the warmer to colder simulations from the top to the bottom panels.
    }
    \label{fig:morphology_plot}
\end{figure*}

\begin{figure}
    \centering
    \resizebox{\hsize}{!}{\includegraphics{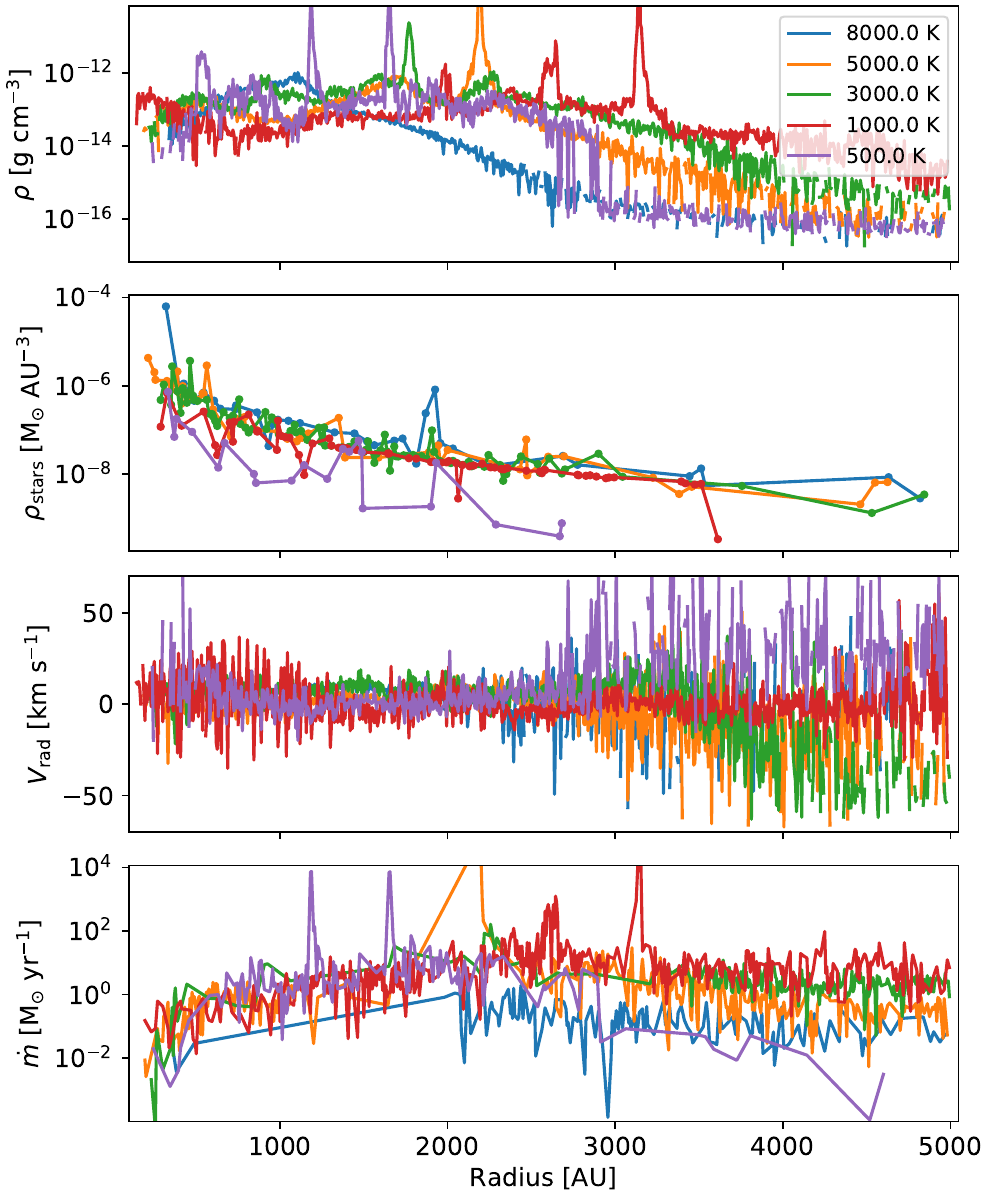}}
    \caption{Radial profiles relative to the MMO for the gas density, stellar mass density without the MMO, radial velocity, and accretion at approximately $ 15000~\mathrm{yr}$, corresponding to the third snapshot of each temperature shown in Fig. \ref{fig:morphology_plot}.
    }
    \label{fig:radial_profiles}
\end{figure}

\subsection{General evolution of the cluster}

The shape of the collapsing gas cloud after the initial collapse is largely sensitive to the cooling and heating processes, and thus by the gas temperature, which is likely to have a significant impact on fragmentation. Fig. \ref{fig:morphology_plot} shows the face-on view of our simulations, where we present the morphology at different times of our simulations with temperatures from $T = 8000 ~\rm K $ to $T = 500 ~ \rm K$ (edge-on view can be found in Appendix \ref{appendix_edge_on}). The maximum gas density reaches $\rho \approx 5.08\times~10^{-6}~\mathrm{g~cm^{-3}}$ for the hottest simulations and $\rho \approx 5.31\times~10^{-3}~\mathrm{g~cm^{-3}}$ for the coldest simulations. In all temperatures we observe a similar initial evolution with a rapid collapse of the central gas cloud, where $25\%$ of the total mass collapsed between $\sim 3000 ~- ~4000 ~ \mathrm{yr}$, thereby producing a larger potential gap in the center of the cluster leading to high accretion rates, runaway collisions and star formation. In nine of our simulations, the MMO was a new sink formed in this central overdense region, whereas in the other run, a sink particle from the initial configuration became the MMO.

After the formation of the MMO (marked as a white star), we observe a comparable overall morphology of the collapsing clouds, maintaining a spherical shape at the beginning and later forming a compact filamentary structure in the center. This elongated overdensity was produced from a disk structure around the MMO, formed as a result of angular momentum conservation.

\begin{table*}
\caption{Summary from our simulations.}
\begin{center}
\begin{tabular}{c c c c c c c c c c c c }
\hline\hline
ID &$\mathrm{Time}$ &$\mathrm{T_{init}}$& $\mathrm{M_{MMO}}$ &$\dot{m}_{\mathrm{MMO}}$ & $\epsilon$ &  $\mathrm{M_{gas}}$ & $\mathrm{T_{mean}}$ & $\mathrm{N_{star}}$ & $\mathrm{M_{star}}$ & $\mathrm{N_{coll}}$ & $\mathrm{M_{esc}}$  \\ 
 $~$&[yr]&[K] & $\mathrm{[M_{\odot}]}$ & $\mathrm{[M_{\odot}~yr^{-1}]}$ & $~$ & $\mathrm{[M_{\odot}]}$ & [K] & $~$& $\mathrm{[M_{\odot}]}$ &$~$ & $\mathrm{[M_{\odot}]}$ \\ \hline
1 &30001.68 & 8000 &18766.54 & 0.0913 & 0.63 & 10102.89 & 8021.57 & 103 & 19711.98 & 1562 & 210.71 \\
2 &30004.78 & 8000& 18109.64 & 0.1835 & 0.60 & 10833.2 & 8016.09 & 116 & 19084.06 & 1401 & 108.23 \\
3&30008.10 & 5000& 21567.15 & 0.2041 & 0.72 & 7897.70 & 5020.12 & 79 & 21973.32 & 2127  & 154.46\\ 
4 &30008.25 & 5000& 21592.66 & 0.1322 & 0.72 & 7409.65 & 5025.25 & 111 & 22444.49 & 2160  & 171.43\\
5 &30008.18 & 3000 & 23216.09 & 0.1550 & 0.77 & 5662.23 & 3036.01 & 102 & 24256.05 & 2691  & 107.21\\
6 &30003.59 & 3000 & 24993.19 & 0.2117 & 0.83 & 3830.09 & 3028.15 & 101 & 26129.79 & 2932  & 65.65\\
7 &28211.18 & 1000 & 25563.54 & 0.5660 & 0.85 & 3874.57 & 1012.28 & 120 & 26105.27 & 1784  & 45.50\\
8 &30002.44 & 1000& 28789.49 & 0.4539 & 0.96 & 1180.65 & 1005.01 & 27 & 28821.57 & 2736 & 23.20\\ 
9 &30009.06 & 500 &28471.44 & 0.6648 & 0.95 & 1529.08 & 506.99 & 35 & 28484.27 & 1678  & 11.72\\ 
10 &30005.51 & 500 & 28875.27 & 0.4751 & 0.96 & 1109.18 & 507.95 & 28 & 28899.14 & 2891  & 17.00\\ \hline
\end{tabular}
\tablefoot{From left to right, the columns correspond to the simulation ID, the final time $\mathrm{Time}$, the final mass of the MMO $\mathrm{M_{MMO}}$, the mean accretion rate of the MMO $\mathrm{\dot{m}_{MMO}}$, the efficiency of the massive object as the mass of the MMO over the total mass of the system $\mathrm{\epsilon}$, the final gas mass $\mathrm{M_{gas}}$, the mean temperature $\mathrm{T_{mean}}$, the final number of protostars $\mathrm{N_{star}}$, the final protostellar mass $\mathrm{M_{star}}$, the total number of collisions $\mathrm{N_{coll}}$, and the total escaped mass $\mathrm{M_{esc}}$.}
\label{tab:result}
\end{center}
\end{table*}

Radial profiles of several quantities related to the general properties of the system are presented in Fig. \ref{fig:radial_profiles}. These profiles correspond to the third snapshots shown in Fig. \ref{fig:morphology_plot}, starting with the gas density in the top panel, stellar mass density excluding the MMO in the second panel, radial velocity of the gas in the third panel, and an analytic accretion rate given by $\dot{m}=4\pi r^2 \rho(r) v_{\rm rad}$ in the fourth panel. In the top panel, we observe a general trend in the shape of the density profiles for the different temperatures. At the center, the gas reaches similar high densities, while at larger radii, they become flatter as the temperature decreases. In the case of $T = 500 ~\rm K$, the density decreases strongly because most of the gas has collapsed. In the second panel, we see that the stellar mass density is higher for warmer temperatures. This is because higher temperatures reduce fragmentation and higher accretion rates from companion stars can be achieved, allowing the formation of more massive companion stars. The third panel in Fig. \ref{fig:radial_profiles} indicates that the gas is not in equilibrium around the MMO, because the gas is collapsing in some regions while expanding in others. Even if some gas evaporates, however, it does not reach the escape velocity of the MMO, keeping the same gas mass or increasing it, if the gas continues to flow toward the MMO. In the bottom panel, we show the radial accretion rate, considering only regions when the radial velocity is toward the center ($V_{\rm rad}<0$). Close to the MMO, the accretion rate drops to approximately $\sim 10^{-2}-1~\mathrm{M_{\odot}~yr^{-1}}$ and remains higher at larger radii for colder temperatures, following the trend of the radial density profile.

\begin{figure}
    \centering
    \resizebox{\hsize}{!}{\includegraphics{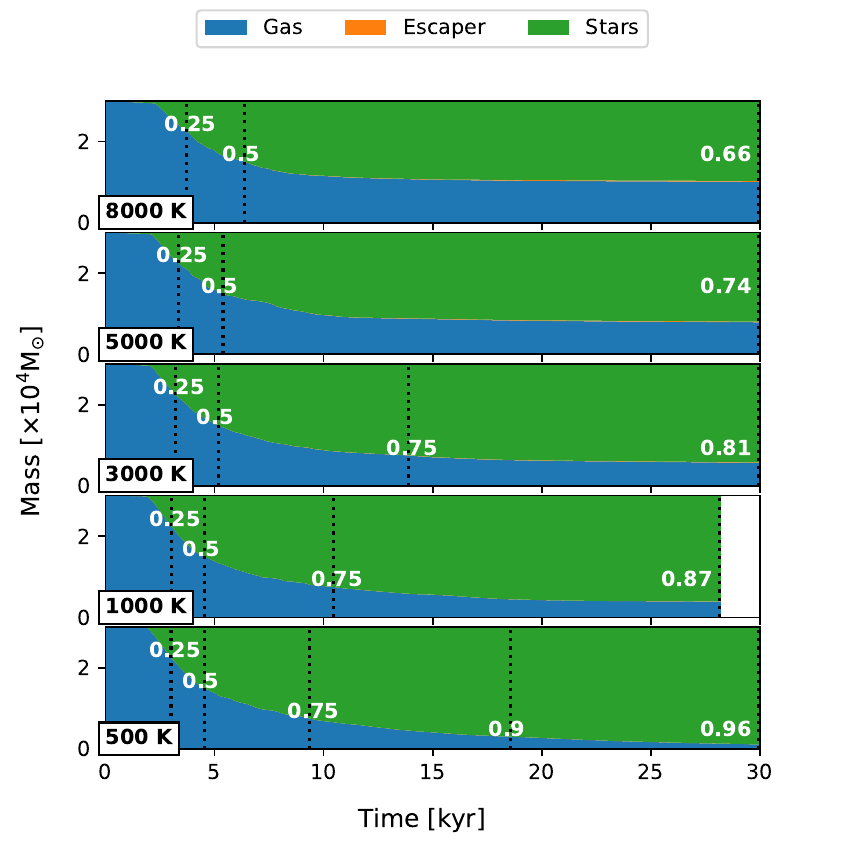}}
    \caption{Time evolution of the mass fraction for the simulations with an initial temperature of $8000~\mathrm{K},~5000~\mathrm{K},~3000~\mathrm{K},~ 1000~\mathrm{K}, \rm and~ 500 ~\mathrm{K}$, corresponding to the gas mass (blue zone), protostars in the cluster (green zone) and the escaped protostellar mass (orange zone). The vertical dotted black lines indicate the times when the protostars accreted 25\%, 50\%, and 75\% of the gas mass.
    }
    \label{fig:mass_all}
\end{figure}

In our simulations, we implicitly assume the presence of metals in the gas, which can cool the gas to temperatures on the order of $100~\mathrm{K}$. \citet{Omukai2008} studied the thermal and chemical evolution of low-metallicity gas clouds exposed to UV radiation. Their one-zone model results indicate that strong UV radiation can suppress $\rm H_2$ formation in a low-metallicity gas cloud, leading to a high-temperature evolution. In the presence of higher metal abundances, the temperature can decrease to $\sim 200~\mathrm{K}$, primarily due to dust thermal emission. Gas evolution with lower temperatures can trigger an early fragmentation of the gas cloud that could generate more sink formation and convert the gas mass into protostellar mass more quickly as we can see in Fig. \ref{fig:mass_all}, where the blue, green and orange regions are the mass from the gas, protostars and escaped protostars of the system. In Fig. \ref{fig:mass_all} we observe a faster reduction of the gas mass for simulations with lower temperatures, with steeper slope and a displacement of the black dashed lines to the left, indicating that more gas mass was accreted at a given time due to gas accretion or star formation. We find in general that the escaped mass from the system can be ignored, since even in the most extreme case the cluster just loses $\sim 0.007 \%$ of the total mass.

In all simulations, an MMO is formed and most of the gas is accreted by the stars by the end of the simulations. In all our simulations we find a similar general behavior that can be divided into three phases: (i) rapid initial contraction, (ii) accelerated formation and predominance of the MMO and (iii) a small cluster with the MMO at its center.

\subsubsection{Initial contraction}

The first phase is marked by a rapid initial contraction of the cluster, triggered by the initial instability with a ratio of $M_{\rm gas}/M_{\rm Jeans} = 10^{0.73}$ for the hottest simulations and $M_{\rm gas}/M_{\rm Jeans} = 10^{2.54}$ for the simulations with an initial temperature of 500 K. These more unstable configurations for lower temperatures provide a favorable scenario for high accretion rates and runaway collisions.

\begin{figure}
    \centering
    \resizebox{\hsize}{!}{\includegraphics{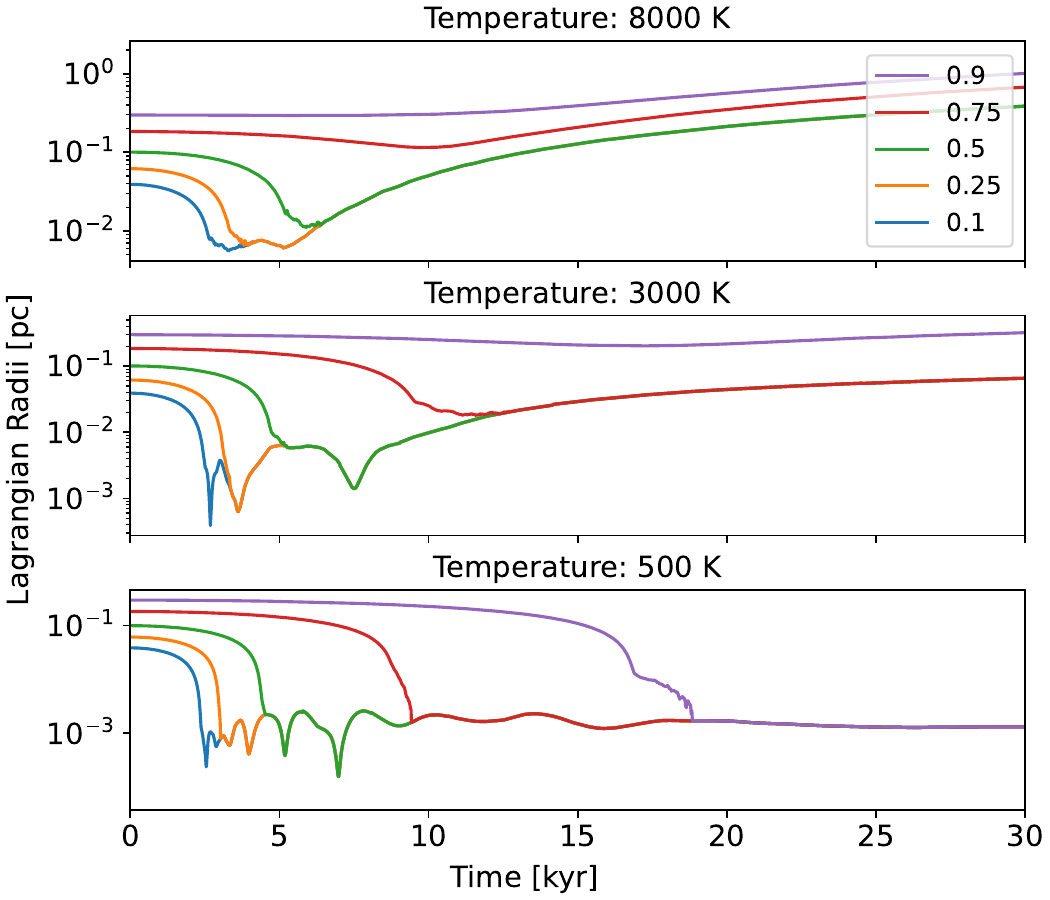}}
    \caption{Time evolution of the 90\%, 75\%, 50\%, 25\%, and 10\% Lagrangian radii for our simulations with temperatures of $8000~\mathrm{K}$, $3000~\mathrm{K}$, and $500~\mathrm{K}$. The panels show the Lagrangian radii considering the total mass of the systems (gas and stars) relative to their centers of mass.}
    \label{fig:LR_8000_2}
\end{figure}

This is illustrated in Fig. \ref{fig:LR_8000_2}, where we present the Lagrangian radii of 10\%, 25\%, 50\%, 75\% and 90\% of the total mass of the system relative to the center of mass for different temperatures. For all temperatures, we observe a rapid core collapse before $\sim 5000 ~\mathrm{yr}$, during which around 50\% of the gas mass was accreted by the stars, as we shown in Fig. \ref{fig:mass_all}. This is caused because the gas cloud departs from equilibrium due to Jeans instability where $M_{\rm gas}>M_{\rm Jeans}$, and the  collapse is isothermal, implying that the pressure forces cannot significantly retard the gas collapse. Considering an initial density $\rho$ as a function of the initial radius $r_0$, we have mass shells within the initial radius $r_0$ with a mean density of $\rho_m(r_0,t_0)$ in a time $t_0$. Assuming that the different shells do not cross each other and as the free-fall timescales as $\rho^{-1/2}$, we will have different free-fall times for different mass shells becoming shorter for the inner regions, since $\rho_m(r,t)$ decreases with increasing $r$. Thus, the inner regions will collapse first, with the other mass shells falling in progressively later. This behavior is best reflected for colder temperatures, where the initial instability is strong enough to collapse most of the gas cloud. In contrast, for warmer temperatures, we observe a higher probability that the MMO may move away from the center of the system, making the accretion less efficient.

\begin{figure*}
    \centering
	\includegraphics[width=17cm]{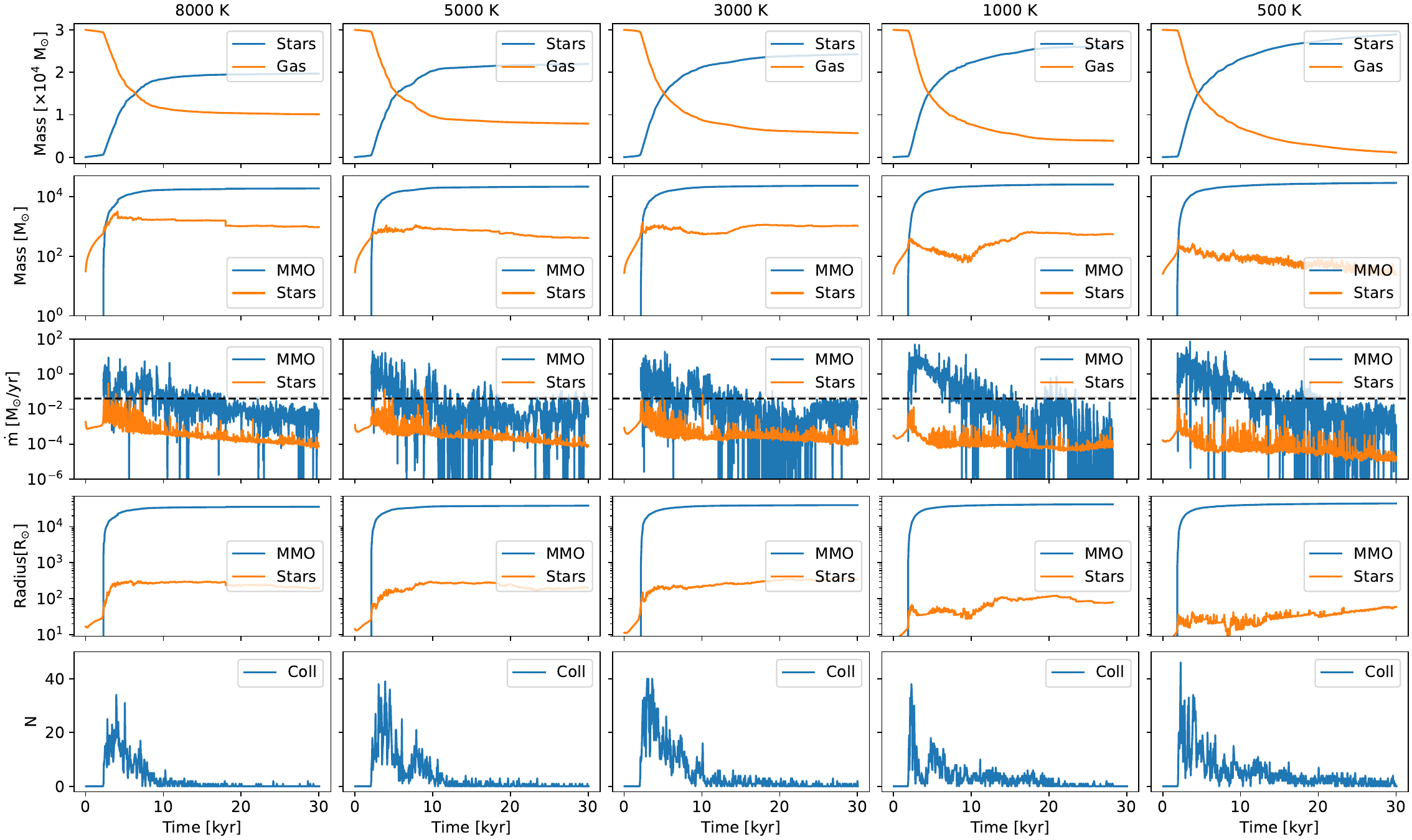}
    \caption{Evolution of the gas and protostellar mass in the simulations with different temperatures (top panel). Comparison of the time evolution of the MMO mass and the total mass of the protostars without the MMO (second panel), accretion rate of the MMO with the mean accretion rate of the protostars without the MMO (third panel), where the black dashed line is the critical accretion rate $\dot{m}_{\rm crit} = 0.04~ \mathrm{M_{\odot
}~yr^{-1}} $ and marks the point where the protostars pass to the SMS Track. We further show the radius of the MMO with the mean radii of the protostars (fourth panel) and the evolution of the total collision rates over time (bottom panel).
    }
    \label{fig:total_datas_all}
\end{figure*}

\subsubsection{Formation of a central object}

After the initial collapse, we observe a high density in the central region of the gas cloud, where the sink forms, which later evolves into the MMO. In nine of our simulations, the MMO was a new sink formed during the first collapse, when the mean density in the warmest simulations within the $25\%$ Lagrangian radius was $\sim 1.11 \times 10^{-10}~\mathrm{g~cm^{-3}}$ and $1.89 \times 10^{-10}~\mathrm{g~cm^{-3}}$ in the coldest simulations. The distance from the center of the system to the position where the MMO was formed is $\sim 0.00348 \pm 0.00166 ~\mathrm{pc}$, so it is in the inner region of the gas cloud with enough gas to accrete and increase in mass.  After the birth of the protostar, it finds itself in a rich environment where it will reach high accretion rates and grows rapidly in mass, from $9.65 \pm 4.13 ~\mathrm{M_{\odot}}$ to $5636.53 \pm 858.14~\mathrm{M_{\odot}}$ in only a thousand years, with a maximum peak in the accretion rate of $20.56 \pm 12.67~\mathrm{M_{\odot}~yr^{-1}}$, due to the accretion of a massive clump formed by jeans instability \citep{hosokawa2016}.

\subsubsection{High accretion rates}\label{second_phase}

While initially the mass of the MMO is smaller than the total mass in the rest of the protostars, eventually the MMO becomes more massive in all the simulations. This transition occurs when the MMO reaches a mass of approximately $645.69 \pm 303.81~\mathrm{M_{\odot}}$. The second phase is marked by very high accretion rates in the MMO, with a mean accretion rate of $1.253 \pm 0.864 ~\mathrm{M_{\odot}~yr^{-1}}$ for warmer simulations and $2.928 \pm 0.271 ~\mathrm{M_{\odot}~yr^{-1}}$ for colder ones. These high accretion rates were reached before $\sim 6000 ~ \mathrm{yr}$, due to the collapse of the gas toward the center of the cluster.

Fig.~\ref{fig:total_datas_all} shows the gas mass fractions and the protostellar mass fractions (top panels), the mass of the MMO and the total mass of the remaining protostars (second panels), accretion rate (third panel), radius of MMO and protostars (fourth panel) and the total collision rate (bottom panels). In the first panels we observe that in the evolution of the total mass of the gas and the protostars, the intersection of the curves occurs at roughly $5219.58 \pm 705.46 ~\mathrm{yr}$, which corresponds to the moment when the dynamics of the system is dominated by the protostars rather than the gas. This point occurs $\sim 1855.27~\mathrm{yr}$ earlier for simulations with $8000~\mathrm{K}$ compared to simulations with $500~\mathrm{K}$. Later we see a difference in the mass exchange between the gas and stars, which is faster for more unstable configurations or colder temperatures. It could be argued that the formation of the MMO is not very sensitive to the initial gas temperature, but the growth could be highly sensitive to the initial instability, driving a faster growth for more unstable configurations. The relatively rapid occurrence of the mass transition also suggests that for our configuration the gas plays a role mostly at the beginning, while subsequently the collision dynamics is more relevant. In the second panels of Fig.~\ref{fig:total_datas_all}, we find the formation of a single massive object (blue line) which consumes most of the available gas mass, while the total mass of the remaining protostars (orange line) reaches a few thousand solar masses. We also note that in most of our simulations, the MMO is a protostar formed due to the gas collapse and not from the protostars that were part of the initial condition. This is because the first component to collapse is the gas and it will concentrate a lot of mass in the center of the cluster, allowing the formation of some new sink particles within a high density region, where the overlap criterion for new sink formation should not be a problem due to the high number of SPH particles in the region.

We observe in the third panels of Fig.~\ref{fig:total_datas_all} that the initial phase of the evolution is marked by high accretion rates of the MMOs, on average with $1.32\pm 3.35~\mathrm{M_\odot ~yr^{-1}}$ at roughly $5000~\mathrm{yr}$ and $1.88\times 10^{-3}\pm 0.01~\mathrm{M_\odot~yr^{-1}}$ for the other protostars at a similar time. The high accretion rates of protostars are caused by the strong gas inflow to the center of the cluster, triggered by the initial gas instability. These high accretion rates rapidly make one of the protostars the MMO in the cluster, leading to a swelling of its radius to $\sim 10^4~ \mathrm{R_{\odot}}$, increasing its probability of collisions very early in the simulation. As a result, the system tends to form a single massive object, with a large mass difference between the MMO and the other protostars in the cluster.

This swelling of the MMO may also explain the high number of collisions observed in the fifth panels of Fig.~\ref{fig:total_datas_all}, where the runaway collision period is, at least partly, a result of the increased radius of the central sink particles and the MMO. The initial high number of collisions is due to the increase in the stellar radius of the MMO, triggered by the initial high accretion rates, which increases its cross section as $\sigma \propto r^2$. The high collision rate is maintained due to the star formation as a result of disk fragmentation around the MMO.

\begin{figure}
    \centering
    \resizebox{\hsize}{!}{\includegraphics{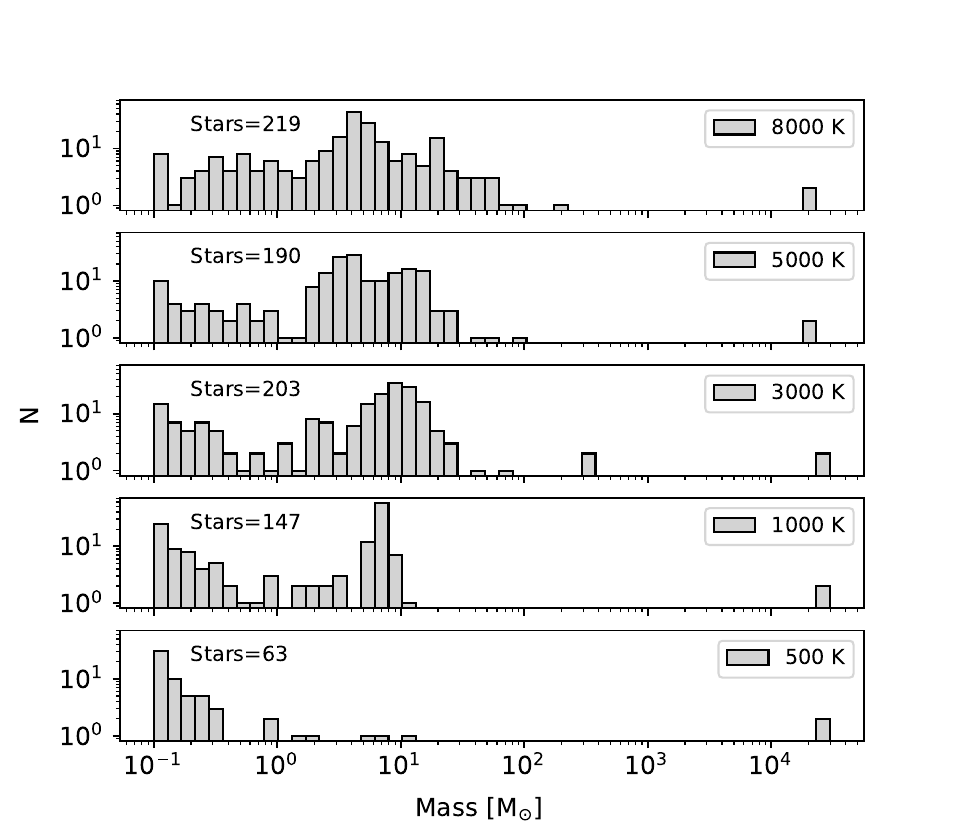}}
    \caption{Combined mass distribution at the end of our simulations for each temperature.}
    \label{fig:mass_ditribution_final}
\end{figure}

\begin{figure}
    \centering
    \resizebox{\hsize}{!}{\includegraphics{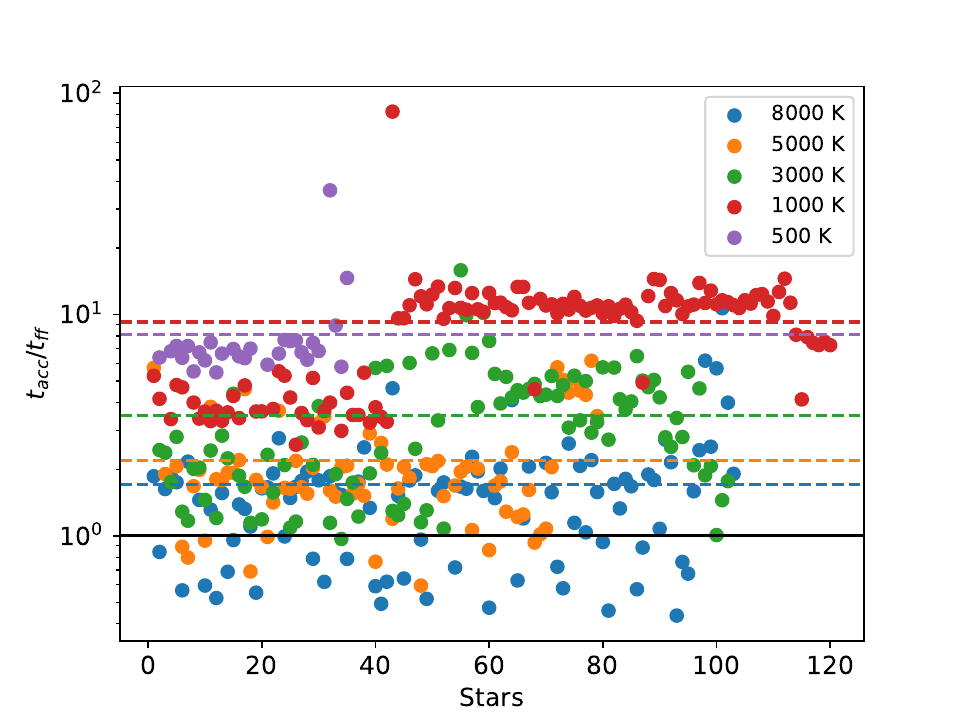}}
    \caption{Mean accretion time over the free-fall time for each star in the cluster at the end of a simulation with each temperature.}
    \label{fig:tacc}
\end{figure}

\subsubsection{MMO and protostellar cluster}

After the growth phase the cluster reaches its final state, where the typical outcome is a small cluster of stars with minimal gas remaining. The typical star masses in the stellar cluster are in the range of $7.35 \pm 20.23 ~\mathrm{M_{\odot}}$, with $81.20 \pm 35.78$ members surrounding the MMO. Up to that time most of the gas was accreted by the stars and the MMO, reaching an efficiency of $\sim 0.62$ for the hottest simulations and $\sim 0.96$ for the coldest one. We have not considered a binary outcome in our results because all second MMOs are less than $400 ~\mathrm{M_{\odot}}$. In Fig. \ref{fig:mass_ditribution_final} we present the combined mass distribution of the stars bound to the MMO in the final state of our simulations. We note that for the simulations with $8000~\mathrm{K}$ there are two trends. The first with a prominent peak that corresponds to stars with masses around $\sim 10~\mathrm{M_{\odot}}$ and with less presence, stars in the lowest limit with $\sim 10^{-1}~\mathrm{M_{\odot}}$. For colder simulations, we observe a reduction of the most massive peak and there are more stars close to the lowest mass peak. The combined final number of stars is also smaller for colder simulations. This is because if we reduce the gas temperature we will have a more unstable configuration, making the gas collapse faster toward the center of the cluster, leaving the stars in a gas-poor environment for accretion. These poor conditions could inhibit the mass growth of protostars, even though they were accreting from the beginning of the gas collapse. When we compare the ratio between the mean accretion time of the protostars that reach the final time with the free-fall time, we find that in the warmest simulation, approximatly $\sim 50\%$ of the stars have $t_{\rm acc}/t_{\rm ff}<1$, and for lower temperatures, $t_{\rm acc}/t_{\rm ff}>>1$ for all final stars. This is shown in Fig. \ref{fig:tacc}, indicating that the gas collapses faster than the protostars could accrete it.

As a result of the reduced temperature, the Jeans mass is also reduced, where $\mathrm{M_{Jeans}} \propto T^{\frac{3}{2}}$. This will produce more fragmentation with smaller masses, reducing the accretion for less massive protostars. Hence, due to the reduced accretion in a gas-poor environment, the stars did not grow large enough in radius to increase their cross section and grow through collisions, so most of the available gas will collapse and be consumed by the MMO.

Our simulations agree with the supercompetitive accretion mechanism proposed by \citet{Chon2020} and not with a fragmentation-induced starvation because $t_{\rm acc}/t_{\rm ff}>>1$, and the gas collapses toward the center of gas cloud, allowing the formation and evolution of a single MMO. 

\begin{figure}
    \centering
    \resizebox{\hsize}{!}{\includegraphics{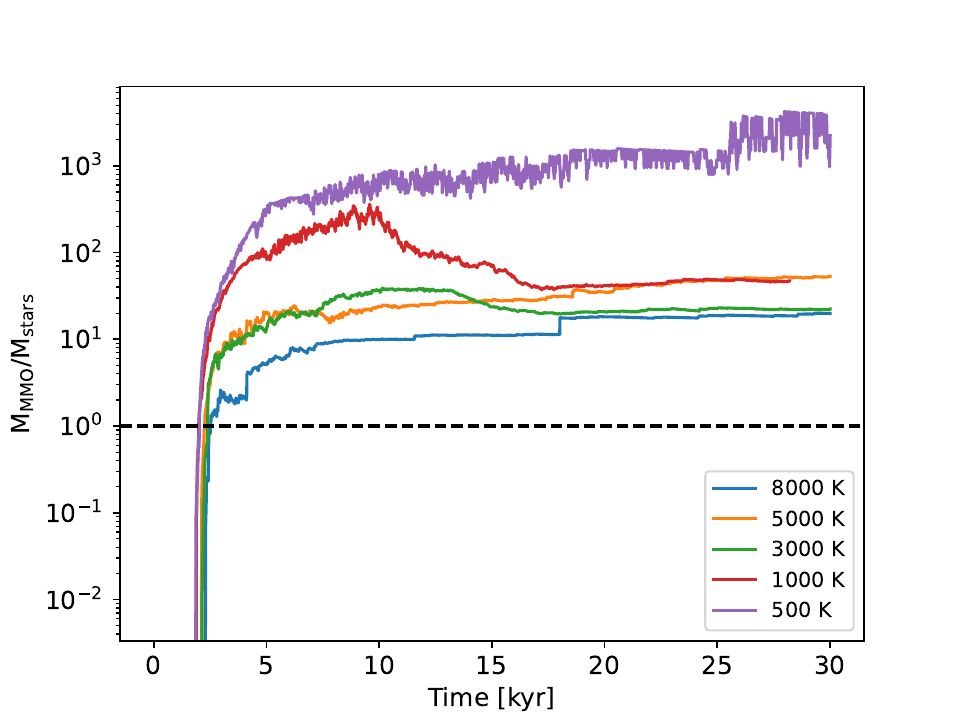}}
    \caption{MMO mass over the total stellar mass without the MMO for each temperature. The horizontal dashed black line shows when $\mathrm{M_{stars}/M_{MMO}} = 1$.}
    \label{fig:mass_stars/mass_mmo}
\end{figure}

In Fig. \ref{fig:mass_stars/mass_mmo} we show the mass of the MMO over the total stellar mass without the MMO, where according to theoretical work by \citet{Dominik_super2023}, the ratio $M_{\rm MMO}/M_{\rm stars}$ must be a factor of $10^{-2}$ to limit the accretion of the MMO. We only see this regime when the MMO was formed, but quickly grows above $1$ and in most of the evolution for all temperatures it is greater than $1$, therefore the stellar mass is not large enough to limit the accretion of the MMO and reach the fragmentation-induced starvation regime.

\begin{figure}
    \centering
    \resizebox{\hsize}{!}{\includegraphics{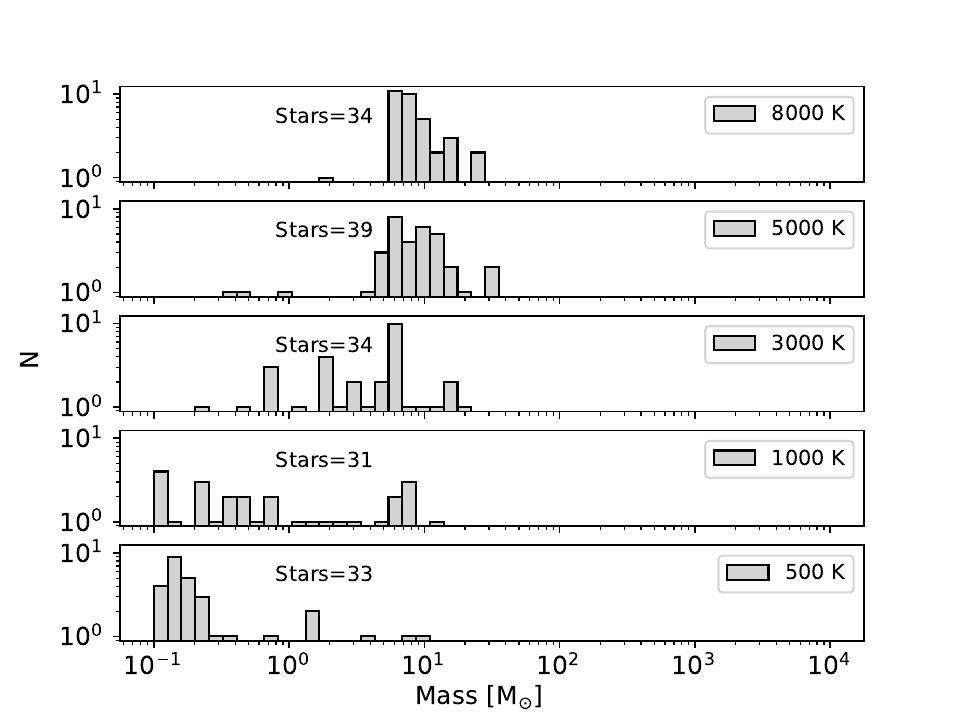}}
    \caption{Combined mass distributions of the escaped stars for simulations with all temperatures.}
    \label{fig:ejected_distribution}
\end{figure}

In Fig. \ref{fig:ejected_distribution} we show the combined mass distribution for all ejected stars at each temperature. For all temperatures we have a similar quantity of ejected stars mainly due to the three-body interactions. The ejected stars show a similar behavior in the peak movement as in Fig. \ref{fig:mass_ditribution_final}, where the peak moves from more massive stars to lighter stars, with $9.38 \pm 4.77~\mathrm{M_{\odot}}$ for the simulations with $8000~\mathrm{K}$ and $0.87 \pm 2.10~\mathrm{M_{\odot}}$ for $500~\mathrm{K}$ simulations.

\subsection{Sink creation mechanism and disk fragmentation}\label{disc}

As we mentioned in the previous section \ref{second_phase}, one possible mechanism for the creation of new sink particles could be the disk fragmentation. In Fig. \ref{fig:morphology_plot} and Fig. \ref{fig:morphology_plot_edge} we observe a disk-like morphology, with a long overdensity around the MMO and both protostars and new sinks are rotating around it. The disk fragmentation implies that the star formation was mainly due to overdensities produced in a turbulent disk structure around the MMO. There are previous numerical studies of disk fragmentation around a massive star in the context of star formation \citep{2020discOliva,2012girichidis,2010peters} and observational evidence of a disk formation around massive stars \citep{2019Maud_observation,2018ginsburg}. Making the disk fragmentation a possible mechanism to explain the high fragmentation and protostar collisions with the MMO. 

The simplest configuration to form a disk structure is a rotating and axisymmetric cloud without magnetic field. Therefore, the gas needs to reach the following conditions:

\begin{itemize}
    \item Axisymmetric gas cloud, with one component of the angular momentum must predominate.
    \item Rotating gas cloud, where the rotational kinetic energy must be higher than the gravitational potential energy.
\end{itemize}

If the gas reaches these conditions, it could form a disk structure around the MMO. In our simulations, we initially impose noncompressive Kolmogorov-turbulence, which induces a net angular momentum of the gas going into gravitational collapse, forming a disk structure around the MMO as a result of angular momentum conservation. After some time, due to the overdensities in the rotating turbulent gas, we observe the formation of sink particles around the MMO.

\begin{figure}
    \centering
    \resizebox{\hsize}{!}{\includegraphics{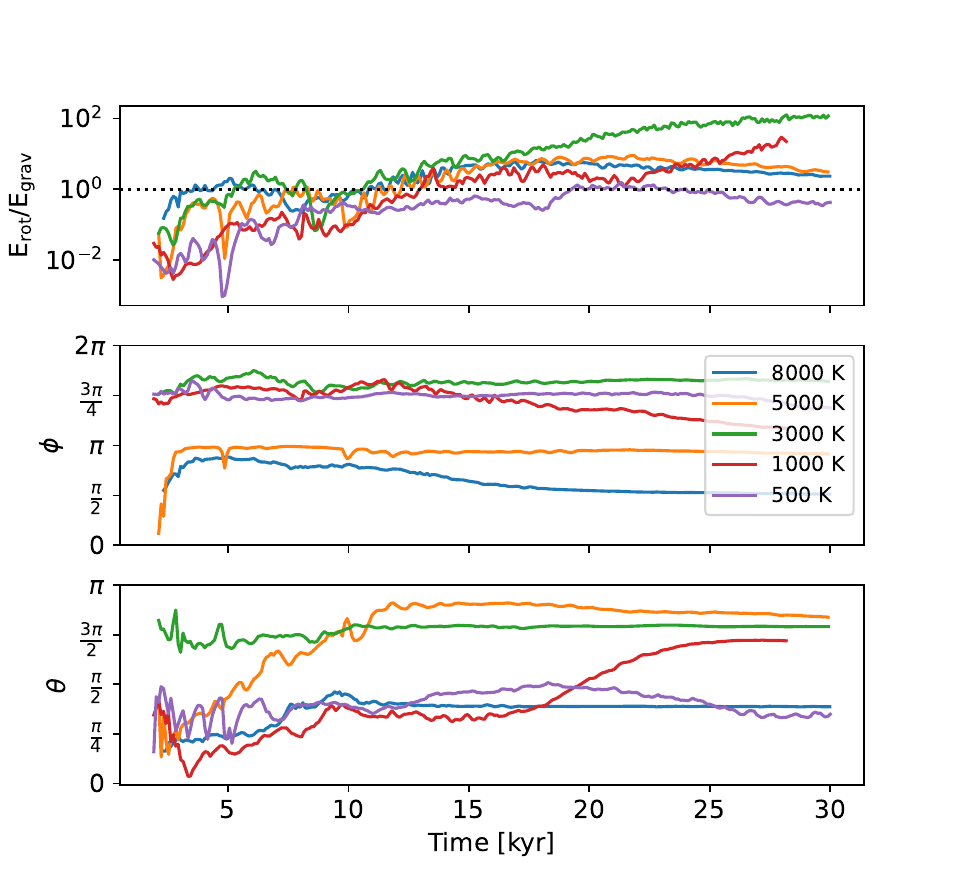}}
    \caption{Ratio of the rotational kinetic energy and potential energy of the gas in a region of $0.01 ~\mathrm{pc}$ around the MMO for each temperature (top panel). The angles $\phi$ and $\theta$ of the total angular momentum are plotted in spherical coordinates (middle and bottom panels).}
    \label{fig:energy_and_L}
\end{figure}

The conditions of axisymmetric and rotational gas cloud are shown in Fig. \ref{fig:energy_and_L}, with the ratio between the rotational kinetic energy and potential energy of the gas in a region of $0.01 ~\mathrm{pc}$ around the MMO and the angular momentum in spherical coordinates. In the top panel we observe that the gas needs some time to increase its rotational energy and to be above the potential energy in order to form a disk structure, needing more time for lower temperatures because the radial velocity toward the center of the gas cloud for colder simulations is higher due to the higher initial instability. For all simulations with temperatures over $500~\mathrm{K}$ the kinetic rotational energy begins to be greater than the potential energy near $\sim 11000~\mathrm{yr}$, and for a temperature of $500~\mathrm{K}$ the rotational energy is above the gravitational energy for a short time, in a time around $\sim 20000~\mathrm{yr}$. Making it more likely that the gas will rotate earlier around the MMO for higher temperatures.

Another necessary condition for disk formation is a predominant angular momentum component. In the middle and bottom panel we show the $\phi$ and $\theta$ angles of the total angular momentum of the gas in a region of $0.01 ~\mathrm{pc}$. In the $\phi$ component we do not observe a lot of variation, being constant most of the time and therefore no azimuthal rotation is observed. On the other side, the $\theta$ component shows more instability for simulations of $1000~\mathrm{K}$ and $500~\mathrm{K}$, needing more time to reach a stable configuration. In general, we observe a predominant direction on time of the total angular momentum of the gas around the MMO.

\begin{table}
\caption{Eigenvalues of the inertia tensor for simulations at each temperature.}
\begin{center}
\begin{tabular}{c c c c }
\hline\hline
$\mathrm{T_{init}}$ [K]& $\lambda_1$ &$\lambda_2$ & $\lambda_3$  \\ 
\hline
8000 & 5.51 & 2.99 & 2.99 \\
5000& 9.17 &4.91 & 4.90\\ 
3000 & 1.31 & 0.69 & 0.65\\
1000 & 4.76 &3.90 & 4.32 \\
500 & 1.15 & 1.39 & 1.63 \\ 
\hline
\end{tabular}
\label{tab:eigenvalues}
\end{center}
\end{table}

We also calculated the components of the inertia tensor relative to the MMO in the following way:

\begin{equation}
    I_{\rm xx} = \sum_{j} m_j(\Delta y_{j,\rm MMO}^2 + \Delta z_{j,\rm MMO}^2)~,
\end{equation}

\begin{equation}
    I_{\rm xy} = \sum_{j} m_j \Delta x_{j,\rm MMO}  \Delta y_{j,\rm MMO}~,
\end{equation}
\\
with $m_j$, the mass of the SPH particle $j$ and $\Delta y_{j,\rm MMO}$ the relative distance in the $y$ component between the gas particle $j$ and the MMO, it is the same for each component of position. The other components of the inertia tensor can be calculated in a similar way. With the inertia tensor we calculate the principal moment of inertia or eigenvalues $\lambda_1$, $\lambda_2$ and $\lambda_3$, which are listed in Table \ref{tab:eigenvalues} for each temperature. We calculate the mean eigenvalues for a time over $10000~\mathrm{yr}$, when approximately the rotational energy is over the gravitational energy. For temperatures of $8000~\mathrm{K},~5000~\mathrm{K}$ and $3000~\mathrm{K}$, the eigenvalues of two components are almost equal and the value for the perpendicular component for each temperature are 1.84, 1.87 and 1.96 times higher than the other two. For a thin disk, the expected value for the perpendicular component is twice the values in the disk plane in agreement with our result for warmer temperatures. Similarly, for temperatures of $1000~\mathrm{K}$ and $500~\mathrm{K}$ the eigenvalues of two components are similar, but the perpendicular component are 1.16 and 1.28 times the values in the disk plane. This difference at colder temperatures could be because the energy and axisymmetrical conditions are reached at later times. Calculating the eigenvalues for later times we note that from $\sim 20000~\mathrm{yr}$ the mean eigenvalues for $1000~\mathrm{K}$ are $\lambda_1=5.60$, $\lambda_2=2.92$ and $\lambda_3=2.89$, for $500~\mathrm{K}$ with a time over $\sim 35000~\mathrm{yr}$, the mean eigenvalues are $\lambda_1=5.61$, $\lambda_2=3.36$ and $\lambda_3=3.56$. This suggests that colder gas clouds need more time to form a structure like a stable thin disk at the same time due to their higher instability.

\begin{figure}
    \centering
    \resizebox{\hsize}{!}{\includegraphics{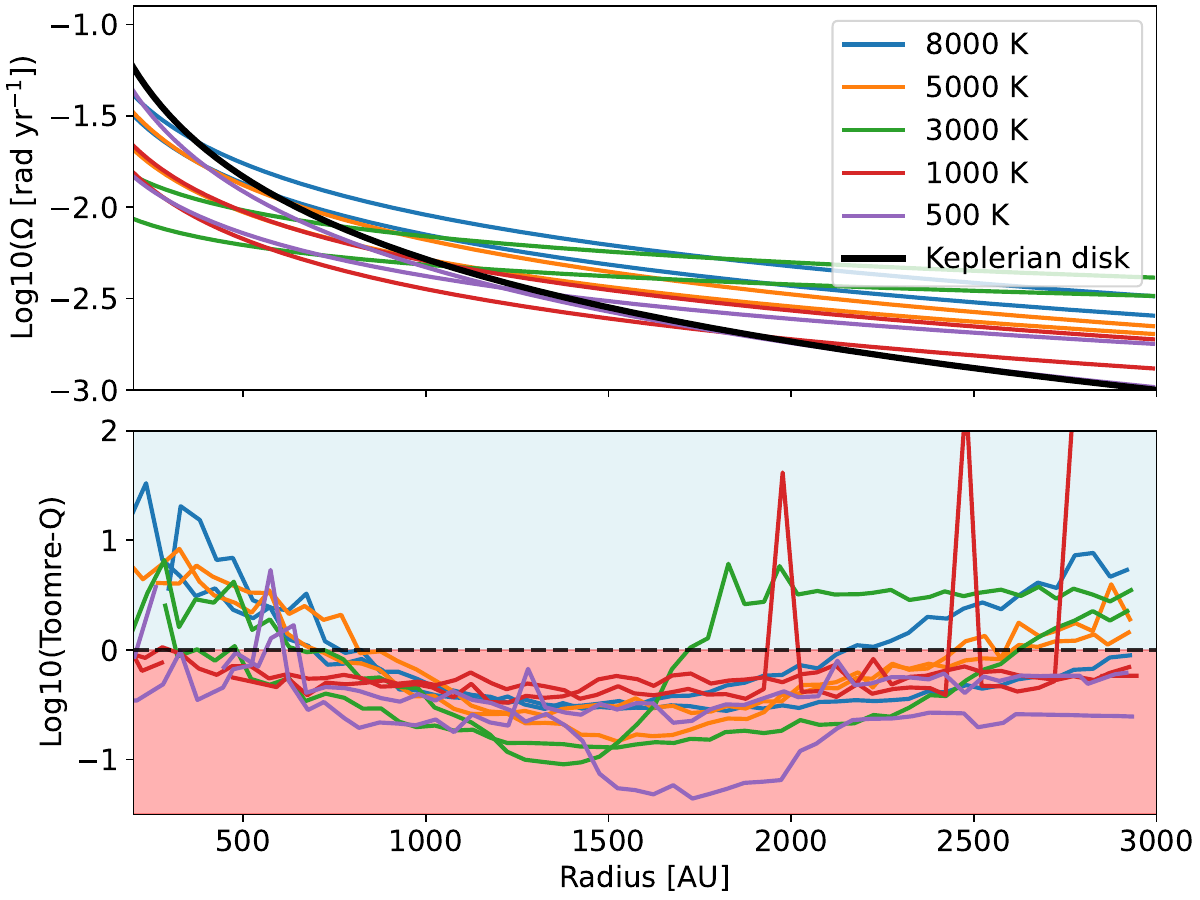}}
    \caption{Radial profiles of the rotation velocity (top panel) and Toomre-Q parameter (bottom panel) around the MMO for all temperatures in the last snapshot shown in Fig \ref{fig:morphology_plot}.}
    \label{fig:Toomre}
\end{figure}

In Fig. \ref{fig:Toomre}, we present the angular velocity and the Toomre-Q parameter as a function of the radius for the last snapshop shown in Fig. \ref{fig:morphology_plot}. In the top panel, the rotational curve for all temperatures tends to be below the Keplerian rotational curve near the MMO, due to the sub-Keplerian correction when considering the contribution of a negative pressure gradient. At larger radii, velocities higher than the Keplerian prediction can be produced by the turbulent behavior triggered by the external torque caused by the stars. In order to quantify gravitational instabilities, we use the Toomre-Q parameter \citep{toomre}, which measures the gravitational stability against small perturbations in a rotating, self-gravitating disc. It is defined as 

\begin{equation}
    Q = \frac{c_s\kappa}{\pi G \sum},
\end{equation}
where $\sum $ is the surface density of the disc, $c_s$ is the sound speed, and $\kappa$ is the epicyclic frequency. For warmer temperatures, the general trend shows a stable disk ($Q > 1$) around the MMO, which then moves into the unstable region at larger radii. In contrast, for simulations with colder temperatures, the disk is Toomre-unstable at all radii, implying higher fragmentation and it can occur anywhere on the disk.

\begin{figure}
    \centering
    \resizebox{\hsize}{!}{\includegraphics{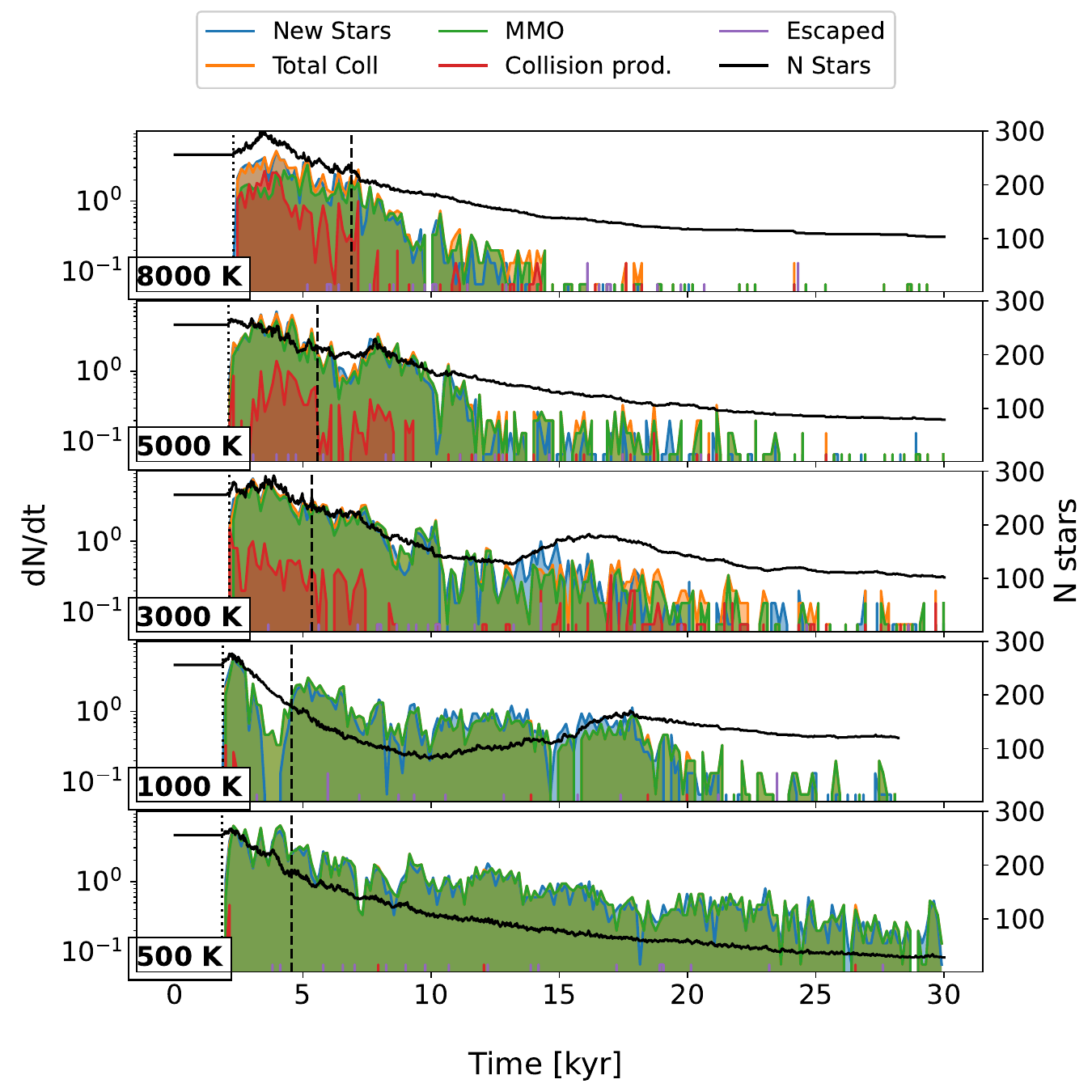}}
    \caption{Time evolution of five rate categories: (1) Formation rate (blue; new stars), (2) total collision rate (orange; total coll.), (3) rate of stars that collided with the MMO (green; MMO), (4) rate of stars that are part of other collision products (red; collision prod.), and (5) rate of stars that escaped from the cluster (purple; escaped). The solid black line shows the number of stars. The vertical dotted and dashed black lines indicate the birth of the MMO and the moment at which the system is dominated by the stellar mass $M_{\rm stars}>M_{\rm gas}$.}
    \label{fig:coll}
\end{figure}

Although we cannot say with certainty that a disk formation exists, the gas tends to rotate around the MMO due to angular momentum conservation, forming a elongated turbulent structure that creates a large number of sink particles that will later collide with the MMO. Most of the protostars that collide with the MMO were formed in this quasi-disk, where the average distance of the created sink particles relative to the MMO for all simulations is $0.0029 \pm 0.0027 ~\mathrm{pc}$, with $1985.90 \pm 514.011$ stars with a distance shorter than $0.01~\mathrm{pc}$, corresponding to $ 98\pm 2\%$ of the created stars. The combined average lifetime of the new sink particles before colliding with the MMO are $207.32\pm 1101.30~\mathrm{yr}$ for a temperature of $8000~\mathrm{K}$, $284.56 \pm 1434.36~\mathrm{yr}$ for a temperature of $5000~\mathrm{K}$, $316.25\pm 1463.48~\mathrm{yr}$ for a temperature of $3000~\mathrm{K}$, $127.80\pm 761.55~\mathrm{yr}$ for a temperature of $1000~\mathrm{K}$, and finally, $47.24\pm 133.44~\mathrm{yr}$ for a temperature of $500~\mathrm{K}$. Colliding the protostars faster in colder gas clouds, having less time to grow in mass.

To understand the relationship between collisions and protostar formation over time, we compare in Fig. \ref{fig:coll} the rates of new protostars (new stars), total collisions (total coll), stars colliding with the MMO (MMO), stars that are part of other collision products (collision prod.) and escaping stars (escaped) and the number of stars in the system (N stars). Our results show that most of the protostars collide with the MMO for colder simulations and the rate of other collision products increases with the temperature. Additionally, most of these other collision products occur in the most chaotic time, between the birth of the MMO (dotted line) and the moment when it becomes the MMO in the system (dashed line), moment when it dominates collisions due to its large cross section.

In our study, we consider that a protostar escaped from the cluster when its distance is greater than $10~R_{\rm gas}$ and its kinetic energy is greater than its potential energy. We observe that in general for all simulations, the number of escaping stars is very small compared to their creation rate. Additionally, the escaped stars are not concentrated in a specific period but are instead distributed throughout the entire evolution. Finally, although there are continuous creations and collisions, the total number of protostars tends to be more or less constant, with a more important decrease in protostars for the coldest simulation.

\subsection{MMO for different temperatures}\label{MMO}

Effects related to the change in gas temperature can greatly affect the evolution of the MMO in a collapsing gas cloud. In this section we will show the main differences in the evolution of the MMO for different gas temperatures.

In all our simulations we reach the formation of a single massive object with a mean mass of $2.10\pm 0.20 \times 10^4~\mathrm{M_{\odot}}$ for the simulations with $T = 8000~\rm K,~5000~\rm K,~3000 ~\rm K$ and $2.70 \pm 0.10 \times 10^4 ~\mathrm{M_{\odot}}$ for the simulations with $T = 1000,~500 ~ \rm K$. As we mentioned in section \ref{second_phase}, the high accretion rates of the central object are driven by a strong gas inflow, triggered by the core contraction. As shown in the third panel of Fig.~\ref{fig:total_datas_all}, in all simulations the MMO accretion rates exceed the critical accretion rate $\dot{m}_{\rm crit} = 0.04~\mathrm{M_{\odot}~yr^{-1}}$ determined by the dashed black line for a long period of time. At the same time, the MMOs enter the SMS track, reaching radii of the order of $ \sim 10^4 ~\mathrm{R_{\odot}}$, and since we assume the most optimistic case analyzed by \citet{Reinoso2023}, with $t_{\mathrm{KH,surf}}=100t_{\mathrm{KH}}$, the SMS track is maintained until the end of the simulation, when only $0.18 \pm 0.11$ of the gas mass remains. Through the swelling and the increase in radius, there is an increase in the cross section of the MMO and, therefore it reaches the highest probability of collision.

\begin{figure}
    \centering
    \resizebox{\hsize}{!}{\includegraphics{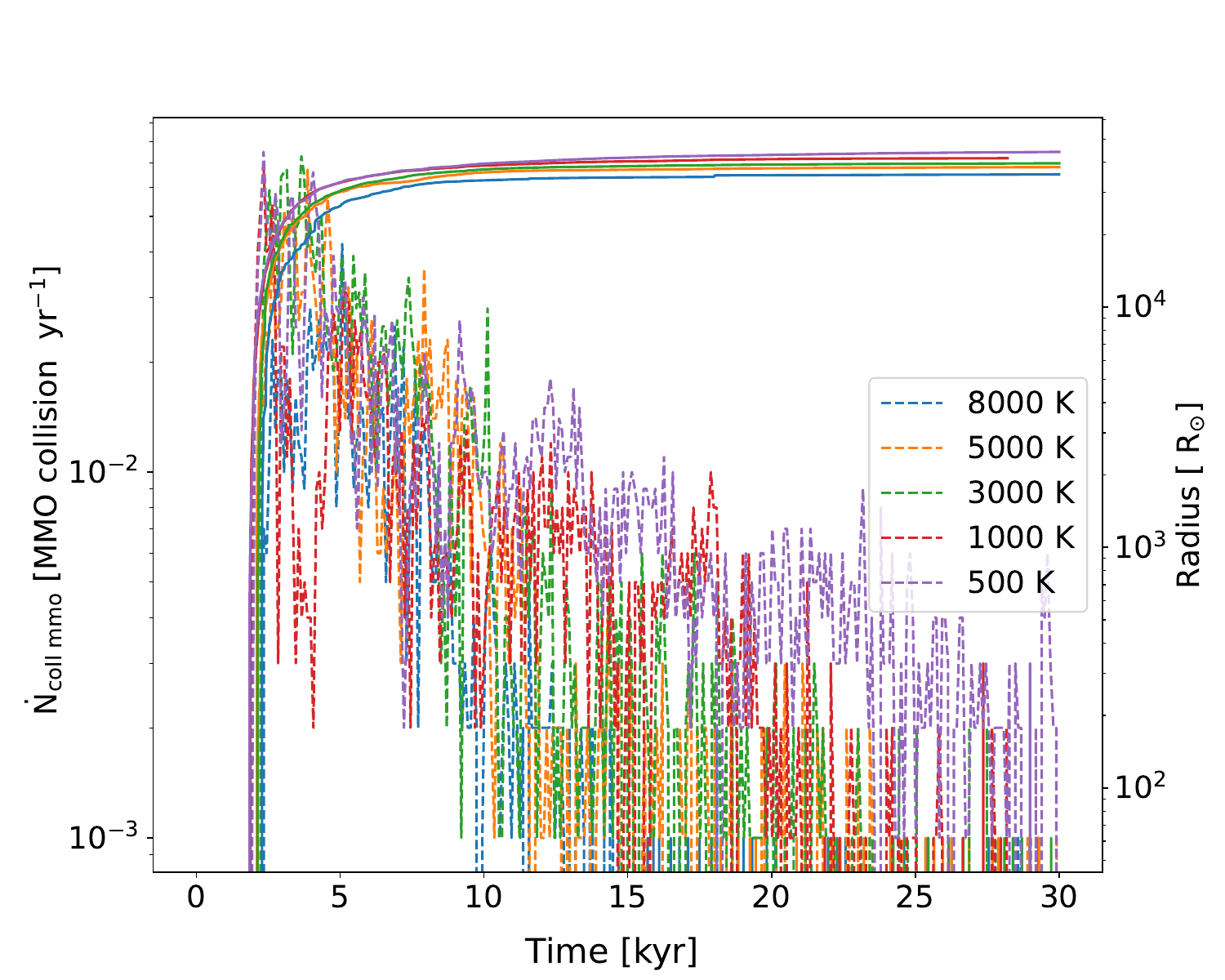}}
    \caption{Evolution of the MMO collision rate (dashed lines) and the radius $R$ of the MMO (solid lines) for each temperature. }
    \label{fig:coll_rate_and_radius}
\end{figure}

In Fig. \ref{fig:coll_rate_and_radius} we show the collision rate and radius of the MMO for a simulation of each temperature. We observe that the highest number of collisions occurs when the MMO increases and reaches the largest stellar radius, leading to a maximum collision rate of $\sim 0.60\pm0.20$ collisions per year.

In colder simulations the accretion rates of the MMO tend to be higher and remain for a longer time above the critical accretion rate. The average accretion rate during this phase, for the simulations with $T = 8000 ~\rm K$, when it exceeds $\dot{m}_{\rm crit}$ is $0.56~\mathrm{M_{\odot}~yr^{-1}}$, and it is maintained for $21671.67~\mathrm{yr}$. In case of $T = 500 ~\rm K$, we have $1.20~\mathrm{M_{\odot}~yr^{-1}}$ for $24106.43~\mathrm{yr}$, making the average accretion rate a factor $2.14$ higher. The high accretion rate triggered by the falling gas leads to a faster formation of a massive object in colder environments.

\begin{figure}
    \centering
    \resizebox{\hsize}{!}{\includegraphics{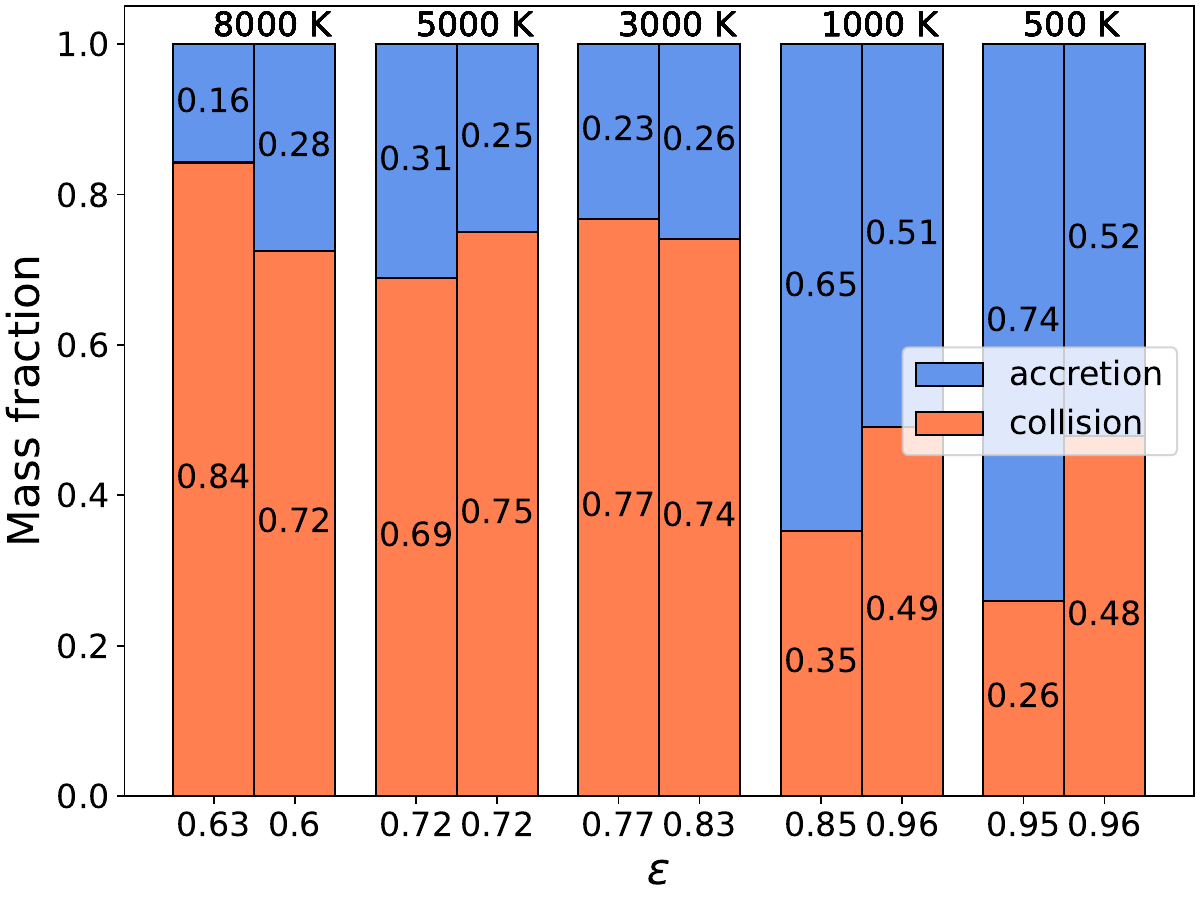}}
    \caption{Final mass fraction of the MMOs for all simulations from $8000~\mathrm{K} - 500 ~\mathrm{K}$ at $\sim 0.3~ \mathrm{Myr}$. We plot the mass acquired through accretion (blue) and the mass obtained via collisions (orange).}
    \label{fig:mass_fraction_mmos}
\end{figure}

The MMO evolves as a SMS due to the high accretion rates and, therefore, the stellar radius leads to an increase of the number of collisions. In our simulations, the only two mechanisms by which MMOs can obtain mass are through gas accretion and/or runaway collisions. In Fig.~\ref{fig:mass_fraction_mmos} we show the mass fraction obtained by collisions and the fraction obtained by accretion for the MMO at the end of the simulations. We clearly note a predominance of the collision mechanism as the main way of obtaining mass for the MMO in simulations with higher temperatures, contributing $75.16 \pm 4.60 ~\%$ of the total mass of the MMO via collisions. On the other hand, the MMOs in the simulations with lower temperatures obtain their mass mainly through accretion, reaching $60.50 \pm 9.55~\%$ through this mechanism.

\begin{figure}
    \centering
    \resizebox{\hsize}{!}{\includegraphics{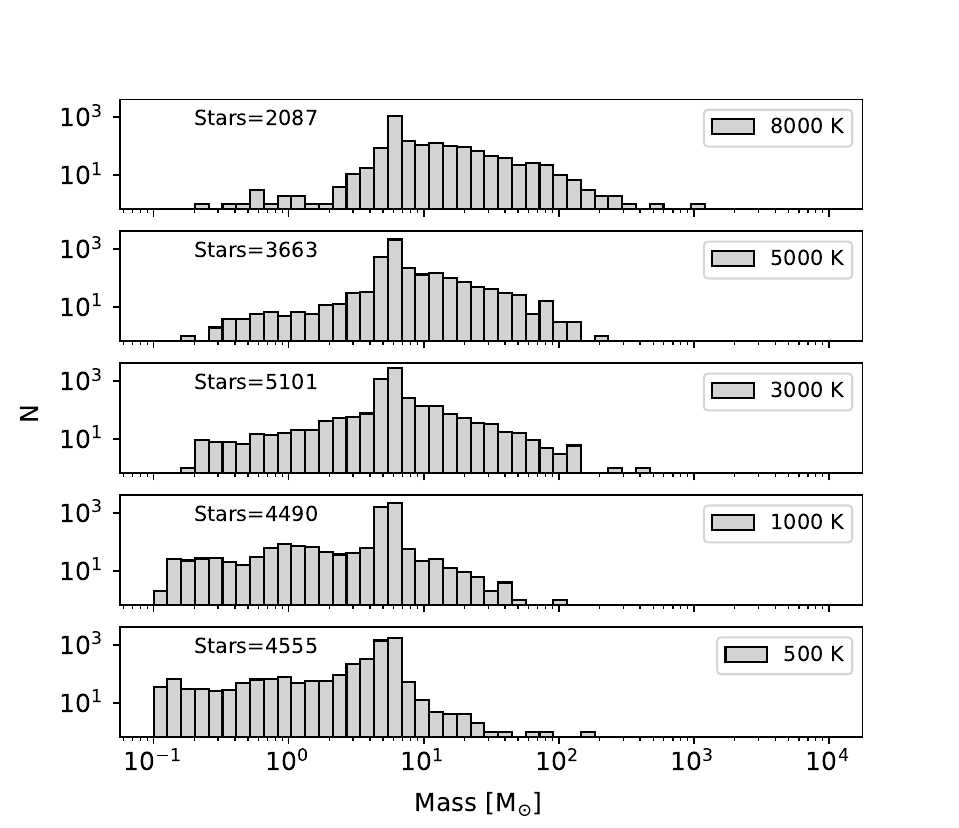}}
    \caption{Combined mass distribution of the stars that collided with the MMO in the simulations with different temperatures.}
    \label{fig:combined_coll_mmo}
\end{figure}

The difference in how the MMO obtains mass in simulations with different temperatures is because the stars that collide with the MMO have more mass in the simulations with higher temperatures, as shown in Fig. \ref{fig:combined_coll_mmo} where we present the combined mass distributions of the stars that collide with the MMO for each temperature. On average the stars that collide with the MMO in the cluster for a temperature of $8000 ~\mathrm{K}$ have more time to grow in mass and radius through accretion compared to those in colder simulations, with average values of $13.9~\mathrm{M_{\odot}},~8.5~\mathrm{M_{\odot}},~7.1~\mathrm{M_{\odot}},~5.2~\mathrm{M_{\odot}},~4.7~\mathrm{M_{\odot}}$ for $8000~\mathrm{K},~5000~\mathrm{K},~3000~\mathrm{K},~1000~\mathrm{K},~500~\mathrm{K}$. The average lifetime of the stars that collide with the MMO for the gas cloud with $8000 ~\mathrm{K}$ is $\sim 1340.37\pm 3835.66 ~\mathrm{yr}$, while it is $\sim 869.86\pm 3141.58 ~\mathrm{yr}$ for the simulation with $500 ~\mathrm{K}$.

\begin{figure}
    \centering
    \resizebox{\hsize}{!}{\includegraphics{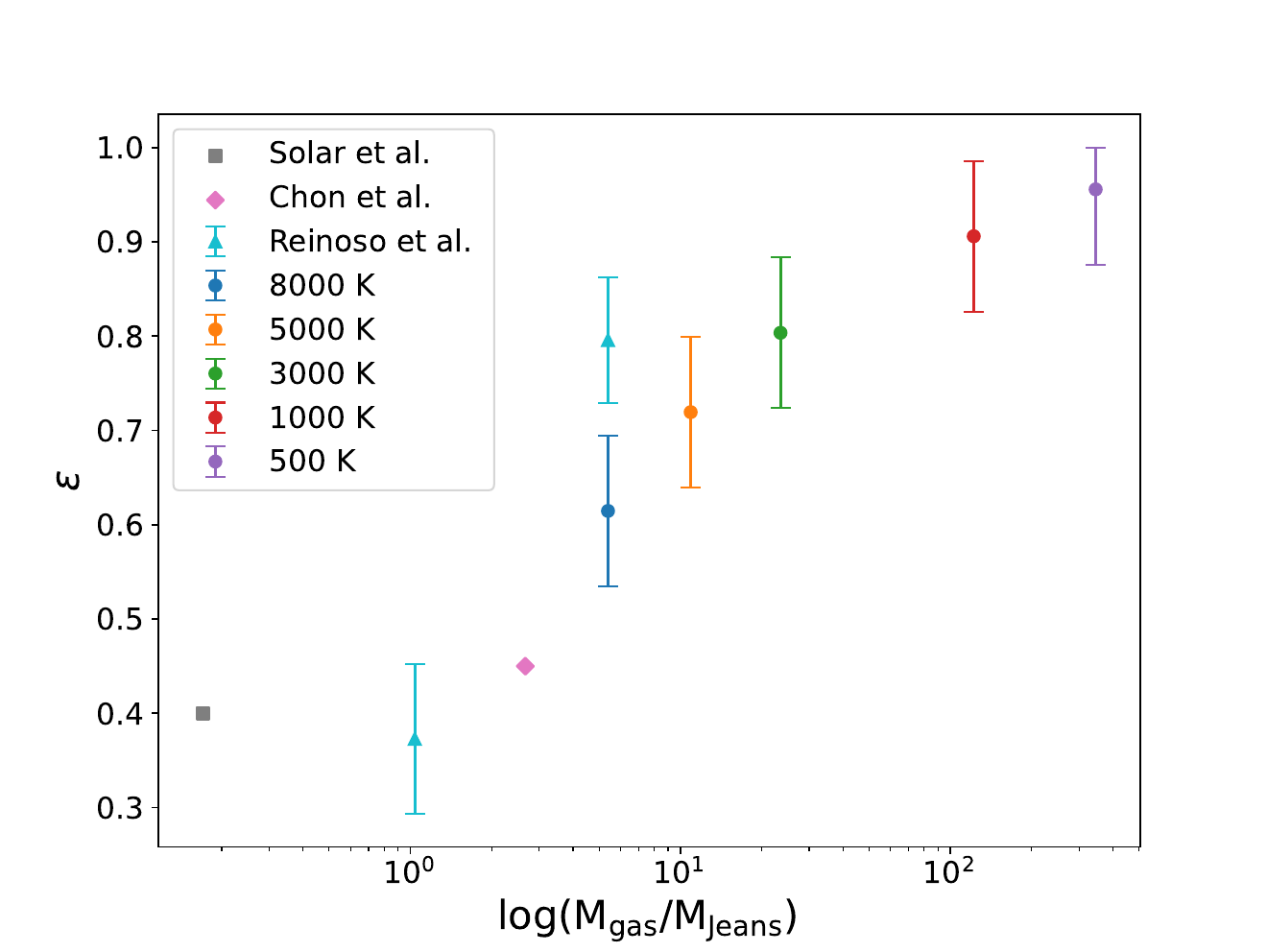}}
    \caption{Efficiency of the MMO formation as a function of the gravitational instability measured via $M_{\rm gas}/M_{\rm Jeans}$ for simulations with temperatures from $8000~\mathrm{K} - 500 ~\mathrm{K}$, including the results presented by \citet{Reinoso2023} (light blue triangles), \citet{paulo2022} (gray square), and \citet{Chon2020} (pink diamond). The error bars were taken into account according to $1\sigma$ error bars by \citet{Reinoso2023}.
    }
    \label{fig:Jeans_all}
\end{figure}

In our simulations we obtain an efficiency ($M_{\rm MMO}/M_{\rm Total}$) for the formation of the MMO of about $0.71\pm 0.08$ in warmer simulations and $0.93\pm 0.04$ in colder ones, which is mainly due to the increased gravitational instability as a result of the high ratio between the gas mass and the Jeans mass. As the collapse of the system is initially driven by the gravitational instability of the gas, it falls toward the center of the cluster faster and higher accretion rates are reached for the central protostars in colder simulations, directly impacting the efficiency of MMO formation, because in less unstable configurations the corresponding timescales will be longer and a fragmentation-induced starvation regime \citep[e.g.,][]{2010peters} could be possible. In Fig. \ref{fig:Jeans_all} we show the efficiencies obtained in simulations with different temperatures as a function of the ratio between the total gas mass and Jeans mass, which we further compare with the results of \citet{Chon2020,paulo2022} and \citet{Reinoso2023}. We observe a correlation between higher efficiencies and more unstable configurations, reaching efficiencies $\gtrsim 80\%$ in the simulations with colder temperatures where the ratio between gas mass and Jeans mass is higher.

\begin{figure}
    \centering
    \resizebox{\hsize}{!}{\includegraphics{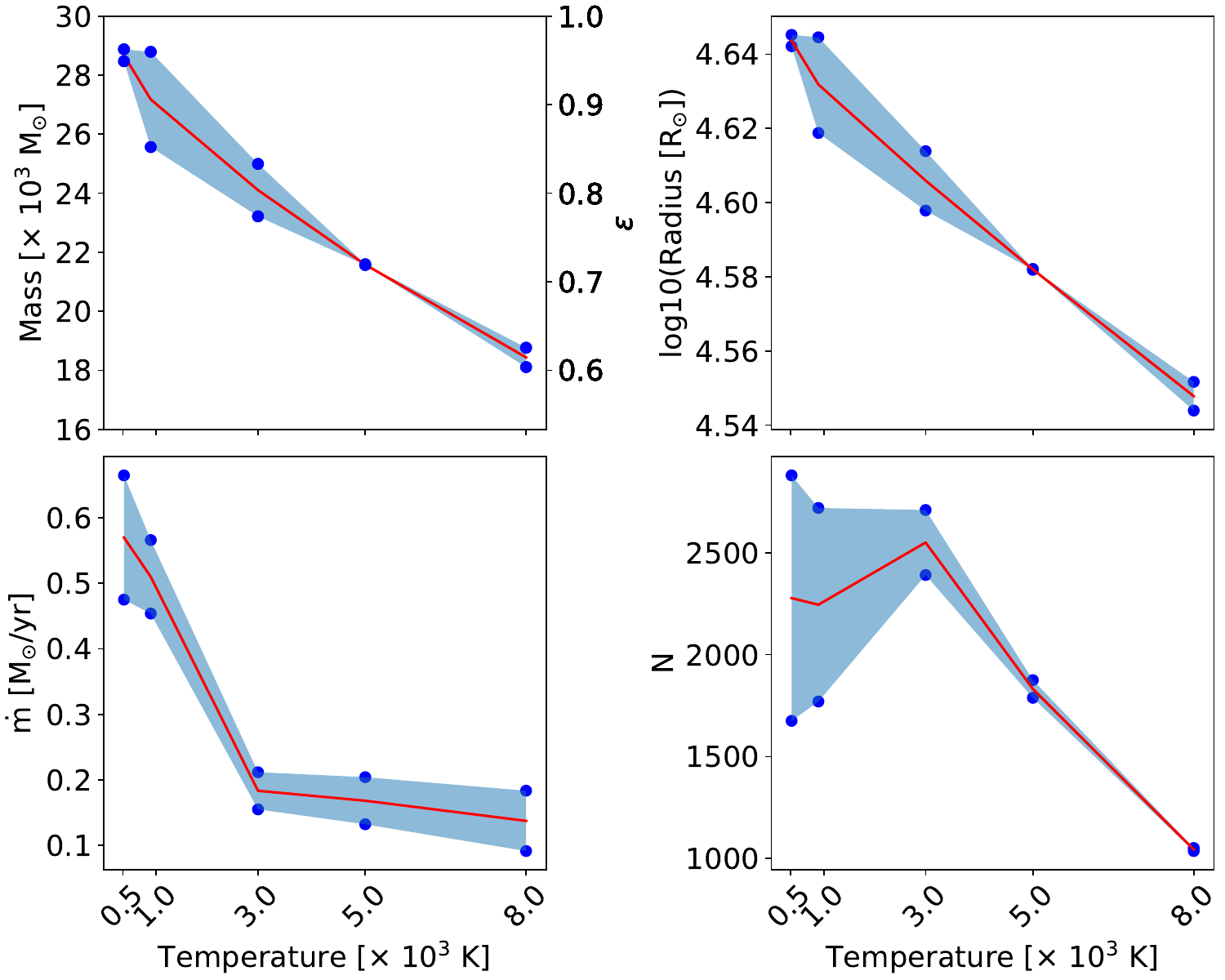}}
    \caption{Final parameters of the MMO (blue dots) as a function of the initial temperature in each simulation. The mass reached at the end of the simulation (upper left), the mean radii (upper right), the mean accretion rates (lower left), and the total number of collisions with the MMO at the end of the simulation (lower right).}
    \label{fig:mmo_final}
\end{figure}

At the end of the simulations, we determine the final mass of the MMO, its radius, mean accretion rate and the total number of collisions with the MMO. The results are presented in Fig.~\ref{fig:mmo_final}. The maximum mass reached by the MMO at the end of each simulation increases for lower temperatures with very high efficiencies due to the high ratio of the total gas mass over Jeans mass, reaching close to $\epsilon \approx 0.95$ for simulations with $500 ~\mathrm{K}$ after $\sim 30000 ~\mathrm{yr}$ and $\epsilon \approx 0.60 $ for simulations with $8000 ~\mathrm{K}$, in agreement with the results reported by \citet{Reinoso2023}, where efficiencies of $\epsilon = 0.90-0.68$ have been reached by evolving the cluster for $200 ~\mathrm{kyr}$.

Observing the radii of the MMOs in the upper right panel of Fig.~\ref{fig:mmo_final}, we notice that they increase monotonically with the mass of the MMO. The mean final radius in simulations of lower temperatures is $\sim 44026.07~ \mathrm{R_{\odot}}$, a factor $\sim 1.25$ higher compared to warmer simulations. After the initial growth of the radius, the MMOs maintain their radius on a similar level due to high accretion rates, independently of their final masses. The mean accretion rate of the MMOs decreases with increasing temperature, averaging $0.14~\mathrm{M_{\odot}~yr^{-1}}$ for the warmest simulations and $0.57~\mathrm{M_{\odot}~yr^{-1}}$ for the coldest one, which could explain the rapid evolution of the MMO in colder simulations. Finally, in the bottom right panel we show the total number of collisions for the different temperatures, with a tendency to have more collisions in configurations with lower temperatures. In colder simulations, we also have a larger variation due to random overdensities leading to disk fragmentation.

\begin{figure*}
    \centering
	\includegraphics[width=17cm]{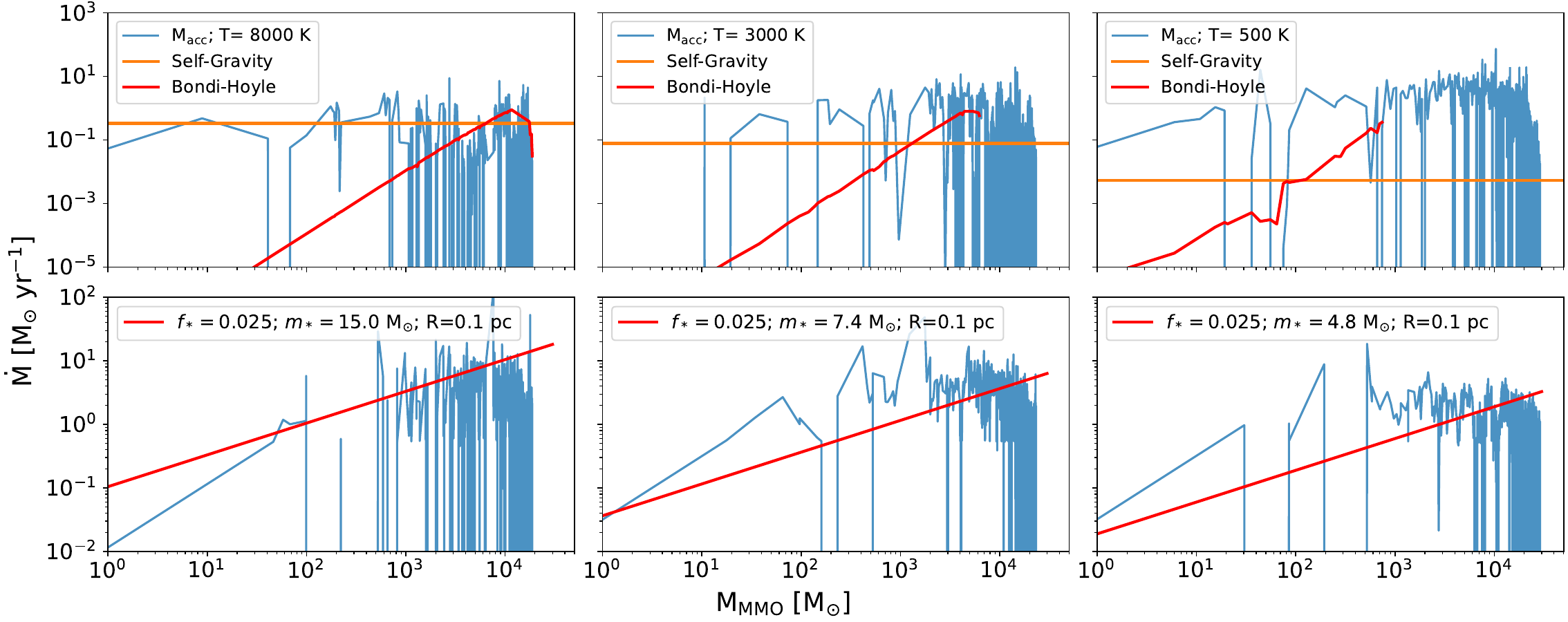}
    \caption{Comparison of the mass growth rate (top panel) and the growth rate due to stellar collisions (bottom panel) as a function of the mass of the MMO for temperatures of $8000 ~\mathrm{K},~3000 ~\mathrm{K}$, and $500 ~\mathrm{K}$ with the theoretical accretion rates of Bondi-Hoyle (red line) and accretion regulated by self-gravity (orange line) from \citet{Dominik_super2023}.}
    \label{fig:bondi_self_coll}
\end{figure*}

\subsection{Accretion regime}

Our results indicate that the supercompetitive accretion scenario can be reached even in the presence of fragmentation. This scenario corresponds to a situation in which only a few central objects grow supermassive with significant accretion inflows of $\gtrsim 0.1~\mathrm{M_{\odot}~yr^{-1}}$, while a larger number of stars compete for the gas reservoir \citep{Chon2020}. The conditions where this scenario develops were studied through analytical arguments by \citet{Dominik_super2023}, taking into account the growth of a massive object via stellar collisions and accretion due to self-gravity and Bondi–Hoyle accretion. They included corrections for angular momentum effects due to a suppression factor \citep{2018inayoshi} and the velocity of the massive objects itself.

They found that at the beginning of the mass growth of the MMO, it will be regulated by self-gravity induced accretion because the density and mass of the central object are sufficiently low, regardless of the number of stars. The accretion rate due to self-gravity depends strongly on the temperature, reaching higher accretion rates for warmer gas clouds, with $\sim 0.2 ~\mathrm{M_{\odot}~yr^{-1}} $ for gas clouds with $8000 ~\mathrm{K}$, and $\sim 10^{-3} ~\mathrm{M_{\odot}~yr^{-1}}$ for temperatures of $300 ~\mathrm{K}$. Only when the central object becomes sufficiently massive the gas accretion can be dominated by Bondi–Hoyle accretion due to the reduction of the mean density in a constant spherical region. Since self-gravity induced accretion depends on temperature as $\propto T^{\frac{3}{2}}$, while Bondi-Hoyle accretion scales as $\propto T^{\frac{-3}{2}}$, colder temperatures allow Bondi–Hoyle accretion to dominate at lower masses of the MMO. Because the cooling and heating processes are complicated, however, there are uncertainties of how the temperature will evolve around the MMO in a realistic scenario.

Calculating the accretion rates of the MMO and the growth via collisions, we can compare our results with the theoretical analysis performed by \citet{Dominik_super2023}. In the top panels in Fig. \ref{fig:bondi_self_coll}, we compare the accretion rates of the MMO for three of our simulations with temperatures of $8000 ~\mathrm{K}$, $3000 ~\mathrm{K}$ and $500 ~\mathrm{K}$, with their analytical predictions for both the accretion rate driven by self-gravity and Bondi–Hoyle accretion rate as a function of the mass of the central massive object. For the self-gravity induced accretion rate we assume an isothermal collapse of $8000 ~\mathrm{K}$, $3000 ~\mathrm{K}$ and $500 ~\mathrm{K}$. While the Bondi–Hoyle accretion rate is formally defined at infinity ($\rho_{\infty}$), we evaluate it through the average density of the cloud. Since Bondi–Hoyle accretion must be evaluated outside the Bondi radius ($R_{\mathrm{Bondi}}$), it is only valid when $R_{\mathrm{Bondi}}<R_{\mathrm{gas}}$.

For a temperature of $8000 ~\mathrm{K}$ and $3000 ~\mathrm{K}$, we observe that the self-gravity inflow generally provides a good approximation for the accretion rate, despite some differences, such as self-gravity underestimates the general accretion flow for $3000 ~\mathrm{K}$ and between $200~\mathrm{M_{\odot}}- 3000~\mathrm{M_{\odot}}$ for $8000 ~\mathrm{K}$. At later times, when the mass of the MMO exceeds $10^4~\mathrm{M_{\odot}}$, it becomes difficult to determine whether accretion is dominated by Bondi–Hoyle or self-gravity accretion in the $8000 ~\mathrm{K}$ simulation. In the case of $3000 ~\mathrm{K}$, since $R_{\mathrm{Bondi}}$ is comparable to $R_{\mathrm{gas}}$, the configuration breaks down the Bondi-Hoyle approximation during the growth of the MMO. In masses larger than $\sim 7000~\mathrm{M_{\odot}}$, self-gravity effects are more relevant. For the simulation of $500 ~\mathrm{K}$, the Bondi-Hoyle accretion can be approximated until masses of $\sim 700~\mathrm{M_{\odot}}$, for the central massive object. For larger masses, the accretion exceeds the Bondi accretion due to it is boosted by the self-gravitational effects of the gas, triggered by the high instability and increased density in the center of the cloud. As a result, the self-gravity drives the gas toward the center of the system, in addition to the gravity of the central massive object. The variability in the accretion rates from our result is also due to fluctuations and substructure within the gas.

In the bottom panels in Fig. \ref{fig:bondi_self_coll}, we compare the growth rate due to stellar collisions with our simulations at three different temperatures inside a region of $0.1~\mathrm{pc}$, where most of the fragments are distributed around the MMO. To calculate the theoretical contribution from stellar collisions, we used the approximation from \citet{Dominik_super2023}, where they consider the average mass of the stars over the timescale for collisions from \citet{Leigh2017}. Under the conditions of supercompetitive accretion, it is possible to assume $m_* \sim \bar{m}$, $R_* \ll R_{\mathrm{MMO}}$ and $R_{\mathrm{cl}} \sim R$, leading to the following equation to estimate the contribution from collisions: 

\begin{equation}
    \dot{M}_{\mathrm{coll}} \sim \frac{12\sqrt{2}R^2_{\mathrm{MMO}}m_*^{3/2}N}{G^3M^{1/2}f_*^3} \left( \frac{G}{2R} \right)^{7/2}\sqrt{M_{\mathrm{tot}}}~,
\end{equation}
where $f_* = M_{\mathrm{cl}}/M_{\mathrm{tot}}$, $M_{\mathrm{cl}}$ is the mass of the cluster, $R_{\mathrm{MMO}}$ is the radius of the MMO, $m_*$ is the average stellar mass, $N$ is the number of stars, $M$ is the mass of the MMO, and $R$ the radius of the cluster. To compare the theoretical estimation to our data, we assume an average mass of the fragments that collide with the MMO of $m_* = 15.0~\mathrm{M_{\odot}}, 7.4~\mathrm{M_{\odot}}$ and $4.8~\mathrm{M_{\odot}}$ for gas clouds with $8000 ~\mathrm{K},3000 ~\mathrm{K}$ and $500 ~\mathrm{K}$. For our simulations, the typical number of stars is $N\approx 145$, with a combined mean mass of $m_{\rm mean}= 5.14~\mathrm{M_{\odot}}$, resulting in a ratio $f_* = 0.024$. In general, we find good agreement between the analytical estimate from \citet{Dominik_super2023} and the mass growth of the MMO due to stellar collisions of our results. As the analytical equation predicts, the contribution from collisions increases with the mass of the MMO, because $R_{\mathrm{MMO}}$ increases with the mass following an SMS stage, leading to a larger cross section. Additionally, the collision contribution is sensitive to the mass of the fragments that collide with the MMO, being more massive for higher temperatures and lower for colder temperatures due to fragmentation effects, where more stars merge with the central massive object but with lower masses. We note that the analytic rate has a better match for the warmest simulations and underestimates the mass growth for colder temperatures when the MMO has a mass lower than $10^4~\mathrm{M_{\odot}}$. The formation of a structure like a disk around the MMO (see section \ref{disc}) will produce uncertainty due to the generation of substructures that later will collapse with the MMO, possibly increasing the collision rates. This could be more relevant for colder temperatures where the gas will produce more fragments and higher star formation rates can be reached. The differences between the analytic equation and our results probably are due to the simplifying assumptions such as an uniform star distribution, a constant number of stars and a constant mass of the fragments within the cluster. Even with the uncertainties associated with the simplifications for the collision rate, we conclude that in general our results are in agreement with the theoretical equations and provide a good approximation to predict the mass of the central massive object obtained via collisions.

\section{Discussion and conclusions}\label{discussion}

When a gas cloud collapses, fragmentation can occur even in the presence of a minimum amount of metals. This leads to the formation of a protostellar cluster that is embedded in a massive gas cloud. Different environmental factors such as the inhibition of $\rm H_2$ formation through ionizing photons from the first generation of stars or a first quasar as a source of UV radiation can influence the evolution of the gas temperature and its collapse behavior. Additionally, metal contamination from halos hosting this first generation of stars can significantly affect its cooling properties \citep{Schneider2006,Schneider2003,Omukai2005}. \citet{Omukai2001} studied the effect of different intensities of far-ultraviolet (FUV) radiation on the primordial star formation and reported that under intense FUV radiation, gas collapse is approximately isothermal and regulated by atomic cooling lines. In contrast, a cloud irradiated with weaker FUV radiation collapses dynamically via $\rm H_2$ cooling and reaches a lower minimum Jeans mass. This leads to increased fragmentation. \citet{Omukai2008} studied the impact of the gas metallicity on the temperature evolution, where gas fragmentation is inevitable above a critical metallicity between $Z_{\rm crit}\approx 5\times 10^{-6}~\rm Z_{\odot} - 3\times 10^{-4}~\rm Z_{\odot}$, and dense clusters of low-mass stars can form. These might still form a remnant intermediate-mass BH through stellar collisions.

We explored the formation of a massive object during the collapse of a protostellar cluster that was embedded in a massive gas cloud following fragmentation, accretion, and collisions on subparsec scales. We investigated the impact of different temperatures on the formation of a massive object and its potential to form a SMBH seed through a simplified equation of state, in which the gas evolved isothermically according to Eq. \ref{eq:EOS}. We performed a series of hydrodynamical and $N$-body simulations using the AMUSE framework, including the routines developed by \citet{Reinoso2023} for the creation of new sink particles, gas accretion, pseudo- (proto-) stellar evolution and stellar collisions. In our simulations, the gas collapsed nearly spherically as a result of the high gravitational instability before it forms an elongated gas cloud around the MMO. Even with this elongated structure, 90\% of the particle formation on subparsec scales occurs close to the MMO and does not prevent the formation of a SMS with $\sim 10^4~\mathrm{M_{\odot}}$. 

The formation of this elongated structure was triggered by the conservation of angular momentum, which led to the formation of a disk and, subsequently, to fragmentation of the gas. Effects of disk fragmentation in massive star formation were explored in previous work and under different conditions \citep{2020discOliva,2012girichidis,2010peters}. The disk was formed from low initial angular momentum in a collapsing gas cloud that grew and formed a disk with spiral arms that fragmented in eccentric and chaotic orbits. \citet{2020discOliva} explored the convergence of their results for different spatial resolutions and reported a minimum resolution for a convergence of the physical properties. Effects such as magnetic fields and supersonic and subsonic turbulence can also affect the homogeneity in the density profile and produce more fragmentation or reduce it. While the fragments are formed, they tend to migrate to the MMO, and most of the sinks in the disk finally merge with the MMO \citep{LatifSchleicher2015,2016sakurai,2019chon}. More work is required to understand whether disk fragmentation is a possible mechanism in the formation of a SMBH seed.

Even with fragmentation, the MMO grows efficiently in mass and reaches masses of $\sim10^4~\mathrm{M_{\odot}}$, which agrees with the results of \citet{Chon2020}, who investigated the impact of metal enrichment in massive gas clouds from a low-metallicity regime of $Z/Z_{\odot} = 10^{-6}$ to a higher metallicity with $Z/Z_{\odot} = 10^{-3}$. They found that for very low and slight metal enrichment, the cloud collapsed and formed a massive object. Even when the cloud presented some fragmentation, most of it either merged with the central star or was ejected from the system. Only when the metallicity was higher than $10^{-3}~Z_{\odot}$ did the central star not become supermassive because the accretion rates were lower. We did not consider the chemical evolution, but employed a highly simplified treatment to investigate the effects of different thermodynamic conditions by varying the gas temperature. We found that in the context of highly Jeans-unstable clouds, a massive object with a mass $\sim 10^4~\mathrm{M_{\odot}}$ can form at different initial temperatures of the gas cloud. Cooling and heating might have an important impact on the evolution of the MMO by decreasing the accretion rate and limiting the growth in mass of the central star, as was shown in the higher-metallicity case by \citet{Chon2020}. For more stable gas clouds, the cooling could lead to the formation of star clusters and might greatly inhibit the accretion of a massive star. We focused on the regime of strong gravitational instability and explored its effects on the formation of a central massive object. We found a correlation between the efficiency in the formation of the MMO and the Jeans instability of the cloud.

Even for higher protostar formation rates, our results agree with the supercompetitive accretion regime proposed by \citet{Chon2020}. We note that most of the created protostars collide with the MMO because they were formed by gravitational instabilities in the quasi-disk structure around the MMO. In our simulations, the protostars collide very quickly with the MMO and do not have enough time to increase in mass and reach the theoretical condition proposed by \citet{Dominik_super2023} that could lead to fragmentation-induced starvation \citep{2010peters} and restrict the accretion of the MMO. The protostars in the cluster have $t_{\rm acc}>>t_{\rm ff}$ on average, so that the gas collapses faster than it can be accreted by the protostars. 

In the supercompetitive accretion scenario, the accretion inflow could be regulated by two mechanisms, self-gravity and Bondi-Hoyle accretion. Our results agree with previous estimates from \citet{Dominik_super2023}, who reported higher accretion rates in earlier states of the MMO and provided a good approximation for accretion rates in gas clouds with higher temperatures. The MMO initially grows rapidly in mass due to high accretion rates driven by self-gravity. At later times, Bondi-Hoyle accretion is boosted by the increasing gas density in the center of the cloud, driven by the self-gravity of the gas. We conclude that in an atomic cooling halo, the theoretical equation provides a good approximation of the accretion rate, which is mainly gas accretion driven by self-gravity.

Additionally, our results showed a lower contribution through collisions for colder temperatures, according to \citet{Dominik_super2023}, because at these temperatures the Jeans mass is lower. This results in less massive fragments, and thus, in a smaller contribution to the MMO. Furthermore, at higher temperatures, the average mass of the fragments that collide with the MMO is higher, which leads to a stronger impact on the growth via collisions. The variability in the growth rate via collisions is due to the large difference in the mass of the protostars that formed in the disk around the MMO.

In the simulations we presented, we did not include mass loss due to collisions. This effect was explored by different authors in different contexts, such as local star clusters \citep{1999sills,2000sills}, and the mass ratios of the collision stars \citep{2002lombardi} and the effect of stellar structure were considered \citep{2008glebbeek}. \citet{Seguel2020} explored the impact of mass loss due to collisions and included both constant mass-loss fractions and analytical models \citep{2002lombardi}. They found that the mass loss can strongly impact the final mass of the SMBH seeds, which in some cases lost between $60\% - 80\%$ of the final mass and $20\% - 40\%$ for the similar context we studied. Nevertheless masses of about $10^4~\mathrm{M_{\odot}}$ were obtained, and this is sufficient for them to be viable candidates for SMBH seeds. These effects might be important for warmer temperatures as well, where the contribution to the final mass of the MMO is $\gtrsim70\%$ and $\gtrsim25\%$ for colder simulations.

We included a recipe for protostellar evolution to evolve the radius with the mass and accretion rate following \citet{Reinoso2023}. This recipe allowed the protostar to expand and contract to the main sequence according to the accretion rate, and this also affected ejections and stellar collisions, which are sensitive to their cross sections. The object with the highest collision probability is the MMO because it has the highest accretion rates, which increases its stellar radius and thereby also its cross section. Due to the high inflow of gas and the growth in radii of the MMO, most of the protostars merge with the MMO, and only some are ejected due to  three-body interactions ($\sim 0.007\%$ of the total mass for the most extreme case). The stellar contraction to the main sequence triggers  UV radiation from the stars, which ionizes the surrounding gas and creates an ionization front. This might affect the accretion rate of the MMO by the surrounding stars that enter the main sequence. Different studies such as \citet{Schleicher2022, Reinoso2023} concluded that the stellar feedback from surrounding stars that enter the main sequence does not appreciably impact the final mass of the MMO. These authors considered a very metal-poor gas environment that collapses approximately isothermally, however, and they did not resolve the chemical evolution of the gas in detail. More work is required to reduce the uncertainty in these estimates and to determine the final masses of the objects that form under these conditions. 

In all our simulations, a central massive object with a mass higher than $\sim10^4~\mathrm{M_{\odot}}$ formed through runaway collisions and high accretion rates. The MMO grew faster in mass and became more massive for colder temperatures through more unstable initial conditions even with vigorous fragmentation. This is similar to previous studies that modeled low-metallicity conditions \citep{Chon2020,paulo2022,Reinoso2023}. For all our simulations, efficiencies higher than $60\%$ were reached and exceeded $90\%$ for colder temperatures due to the strong gravitational instability. We observed a trend for the formation of a massive central object even with some fragmentation when the gas cloud was sufficiently unstable. The mechanism by which the MMO grows in mass is primarily driven by collisions at higher temperatures, while accretion becomes the dominant process in simulations with lower temperatures. This is a result of the combination of the timescale for the formation of the quasi-disk structure, which is the primary mechanism for the creation of new sinks in our simulations, and the lower Jeans mass at lower temperatures. The typical final outcome in our simulations is a central SMS with a small cluster of stars with masses of about $0.1-100~\mathrm{M_{\odot}}$ for warmer simulations and $0.1-1~\mathrm{M_{\odot}}$ for colder temperatures.

%\section*{Acknowledgements}
\begin{acknowledgements}
PS acknowledges support through ANID/Doctorado en el Extranjero convocatoria 2022 (funding number 72220198), the Federal Ministry of Education and Research and the state governments for supporting this project as part of the joint funding of National High Performance Computing (NHR) and the Kultrun cluster hosted at the Departamento de Astronomía, Universidad de Concepción. We gratefully acknowledge support by the ANID BASAL project FB21003, as well as via Fondecyt Regular (project code 1201280) and ANID QUIMAL220002. BR acknowledges funding through ANID (CONICYT-PFCHA/Doctorado acuerdo bilateral DAAD/62180013), DAAD (funding program number 57451854), and the International Max Planck Research School for Astronomy and Cosmic Physics at the University of Heidelberg (IMPRS-HD). BR acknowledges the support by the European Research Council via ERC Consolidator grant KETJU (no. 818930). DRGS thanks for funding via the  Alexander von Humboldt - Foundation, Bonn, Germany. RSK acknowledges financial support from the European Research Council via the ERC Synergy Grant ``ECOGAL'' (project ID 855130),  from the German Excellence Strategy via the Heidelberg Cluster of Excellence (EXC 2181 - 390900948) ``STRUCTURES'', and from the German Ministry for Economic Affairs and Climate Action in project ``MAINN'' (funding ID 50OO2206). RSK also thanks for computing resources provided by the Ministry of Science, Research and the Arts (MWK) of the State of Baden-W\"{u}rttemberg through bwHPC and the German Science Foundation (DFG) through grants INST 35/1134-1 FUGG and 35/1597-1 FUGG, and also for data storage at SDS@hd funded through grants INST 35/1314-1 FUGG and INST 35/1503-1 FUGG. RB acknowledges support by the Deutsche Forschungsgemeinschaft (DFG, German Research Foundation) under Germany’s Excellence Strategy – EXC 2121 "Quantum Universe" – 390833306.
\end{acknowledgements}

\bibliographystyle{aa}
\bibliography{astro}

\begin{thebibliography}{102}
\expandafter\ifx\csname natexlab\endcsname\relax\def\natexlab#1{#1}\fi

\bibitem[{{Abel} {et~al.}(2002){Abel}, {Bryan}, \& {Norman}}]{Abel2002}
{Abel}, T., {Bryan}, G.~L., \& {Norman}, M.~L. 2002, Science, 295, 93

\bibitem[{{Alister Seguel} {et~al.}(2020){Alister Seguel}, {Schleicher}, {Boekholt}, {Fellhauer}, \& {Klessen}}]{Seguel2020}
{Alister Seguel}, P.~J., {Schleicher}, D.~R.~G., {Boekholt}, T.~C.~N., {Fellhauer}, M., \& {Klessen}, R.~S. 2020, \mnras, 493, 2352

\bibitem[{{Ba{\~n}ados} {et~al.}(2021){Ba{\~n}ados}, {Mazzucchelli}, {Momjian}, {Eilers}, {Wang}, {Schindler}, {Connor}, {Andika}, {Barth}, {Carilli}, {Davies}, {Decarli}, {Fan}, {Farina}, {Hennawi}, {Pensabene}, {Stern}, {Venemans}, {Wenzl}, \& {Yang}}]{Banados2021}
{Ba{\~n}ados}, E., {Mazzucchelli}, C., {Momjian}, E., {et~al.} 2021, \apj, 909, 80

\bibitem[{{Ba{\~n}ados} {et~al.}(2018){Ba{\~n}ados}, {Venemans}, {Mazzucchelli}, {Farina}, {Walter}, {Wang}, {Decarli}, {Stern}, {Fan}, {Davies}, {Hennawi}, {Simcoe}, {Turner}, {Rix}, {Yang}, {Kelson}, {Rudie}, \& {Winters}}]{Banados2018}
{Ba{\~n}ados}, E., {Venemans}, B.~P., {Mazzucchelli}, C., {et~al.} 2018, \nat, 553, 473

\bibitem[{{Becerra} {et~al.}(2015){Becerra}, {Greif}, {Springel}, \& {Hernquist}}]{Becerra2015}
{Becerra}, F., {Greif}, T.~H., {Springel}, V., \& {Hernquist}, L.~E. 2015, \mnras, 446, 2380

\bibitem[{{Begelman} \& {Shlosman}(2009)}]{Begelman2009}
{Begelman}, M.~C. \& {Shlosman}, I. 2009, \apjl, 702, L5

\bibitem[{{Benz}(1990)}]{Benz1990}
{Benz}, W. 1990, in Numerical Modelling of Nonlinear Stellar Pulsations Problems and Prospects, ed. J.~R. {Buchler}, 269

\bibitem[{{Boekholt} {et~al.}(2018){Boekholt}, {Schleicher}, {Fellhauer}, {Klessen}, {Reinoso}, {Stutz}, \& {Haemmerl{\'e}}}]{Boekholt2018}
{Boekholt}, T.~C.~N., {Schleicher}, D.~R.~G., {Fellhauer}, M., {et~al.} 2018, \mnras, 476, 366

\bibitem[{{Bogd{\'a}n} {et~al.}(2024){Bogd{\'a}n}, {Goulding}, {Natarajan}, {Kov{\'a}cs}, {Tremblay}, {Chadayammuri}, {Volonteri}, {Kraft}, {Forman}, {Jones}, {Churazov}, \& {Zhuravleva}}]{2024uhz1}
{Bogd{\'a}n}, {\'A}., {Goulding}, A.~D., {Natarajan}, P., {et~al.} 2024, Nature Astronomy, 8, 126

\bibitem[{{Bromm} \& {Loeb}(2003)}]{Bromm2003}
{Bromm}, V. \& {Loeb}, A. 2003, \apj, 596, 34

\bibitem[{{Chon} \& {Hosokawa}(2019)}]{2019chon}
{Chon}, S. \& {Hosokawa}, T. 2019, \mnras, 488, 2658

\bibitem[{{Chon} \& {Omukai}(2020)}]{Chon2020}
{Chon}, S. \& {Omukai}, K. 2020, \mnras, 494, 2851

\bibitem[{{Clark} {et~al.}(2011){Clark}, {Glover}, {Smith}, {Greif}, {Klessen}, \& {Bromm}}]{Clark2011}
{Clark}, P.~C., {Glover}, S. C.~O., {Smith}, R.~J., {et~al.} 2011, Science, 331, 1040

\bibitem[{{Das} {et~al.}(2021){Das}, {Schleicher}, {Leigh}, \& {Boekholt}}]{Das2021}
{Das}, A., {Schleicher}, D. R.~G., {Leigh}, N. W.~C., \& {Boekholt}, T. C.~N. 2021, \mnras, 503, 1051

\bibitem[{{Davies} {et~al.}(2011){Davies}, {Miller}, \& {Bellovary}}]{Davies11}
{Davies}, M.~B., {Miller}, M.~C., \& {Bellovary}, J.~M. 2011, \apjl, 740, L42

\bibitem[{{Devecchi} \& {Volonteri}(2009)}]{Devecchi2009}
{Devecchi}, B. \& {Volonteri}, M. 2009, \apj, 694, 302

\bibitem[{{Escala}(2021)}]{Escala2021}
{Escala}, A. 2021, \apj, 908, 57

\bibitem[{{Fan} {et~al.}(2006){Fan}, {Strauss}, {Richards}, {Hennawi}, {Becker}, {White}, {Diamond-Stanic}, {Donley}, {Jiang}, {Kim}, {Vestergaard}, {Young}, {Gunn}, {Lupton}, {Knapp}, {Schneider}, {Brandt}, {Bahcall}, {Barentine}, {Brinkmann}, {Brewington}, {Fukugita}, {Harvanek}, {Kleinman}, {Krzesinski}, {Long}, {Neilsen}, {Nitta}, {Snedden}, \& {Voges}}]{Fan2006}
{Fan}, X., {Strauss}, M.~A., {Richards}, G.~T., {et~al.} 2006, \aj, 131, 1203

\bibitem[{{Fraser} {et~al.}(2017){Fraser}, {Casey}, {Gilmore}, {Heger}, \& {Chan}}]{Fraser2017}
{Fraser}, M., {Casey}, A.~R., {Gilmore}, G., {Heger}, A., \& {Chan}, C. 2017, \mnras, 468, 418

\bibitem[{{Fujii} {et~al.}(2007){Fujii}, {Iwasawa}, {Funato}, \& {Makino}}]{Fujii2007}
{Fujii}, M., {Iwasawa}, M., {Funato}, Y., \& {Makino}, J. 2007, \pasj, 59, 1095

\bibitem[{{Gerritsen} \& {Icke}(1997)}]{Gerritsen1997}
{Gerritsen}, J.~P.~E. \& {Icke}, V. 1997, \aap, 325, 972

\bibitem[{{Ginsburg} {et~al.}(2018){Ginsburg}, {Bally}, {Goddi}, {Plambeck}, \& {Wright}}]{2018ginsburg}
{Ginsburg}, A., {Bally}, J., {Goddi}, C., {Plambeck}, R., \& {Wright}, M. 2018, \apj, 860, 119

\bibitem[{{Girichidis} {et~al.}(2012){Girichidis}, {Federrath}, {Banerjee}, \& {Klessen}}]{2012girichidis}
{Girichidis}, P., {Federrath}, C., {Banerjee}, R., \& {Klessen}, R.~S. 2012, \mnras, 420, 613

\bibitem[{{Glebbeek} {et~al.}(2008){Glebbeek}, {Pols}, \& {Hurley}}]{2008glebbeek}
{Glebbeek}, E., {Pols}, O.~R., \& {Hurley}, J.~R. 2008, \aap, 488, 1007

\bibitem[{{Greif} {et~al.}(2011){Greif}, {Springel}, {White}, {Glover}, {Clark}, {Smith}, {Klessen}, \& {Bromm}}]{Greif2011}
{Greif}, T.~H., {Springel}, V., {White}, S. D.~M., {et~al.} 2011, \apj, 737, 75

\bibitem[{{Grete} {et~al.}(2019){Grete}, {Latif}, {Schleicher}, \& {Schmidt}}]{Grete2019}
{Grete}, P., {Latif}, M.~A., {Schleicher}, D.~R.~G., \& {Schmidt}, W. 2019, \mnras, 487, 4525

\bibitem[{{Habouzit} {et~al.}(2016){Habouzit}, {Volonteri}, {Latif}, {Dubois}, \& {Peirani}}]{2016Habouzit}
{Habouzit}, M., {Volonteri}, M., {Latif}, M., {Dubois}, Y., \& {Peirani}, S. 2016, \mnras, 463, 529

\bibitem[{{Haemmerl{\'e}} {et~al.}(2021){Haemmerl{\'e}}, {Klessen}, {Mayer}, \& {Zwick}}]{Haemmerle2021}
{Haemmerl{\'e}}, L., {Klessen}, R.~S., {Mayer}, L., \& {Zwick}, L. 2021, \aap, 652, L7

\bibitem[{{Haemmerl{\'e}} {et~al.}(2019){Haemmerl{\'e}}, {Meynet}, {Mayer}, {Klessen}, {Woods}, \& {Heger}}]{Haemmerle2019}
{Haemmerl{\'e}}, L., {Meynet}, G., {Mayer}, L., {et~al.} 2019, \aap, 632, L2

\bibitem[{{Haemmerl{\'e}} {et~al.}(2018){Haemmerl{\'e}}, {Woods}, {Klessen}, {Heger}, \& {Whalen}}]{Haemmerle2018}
{Haemmerl{\'e}}, L., {Woods}, T.~E., {Klessen}, R.~S., {Heger}, A., \& {Whalen}, D.~J. 2018, \mnras, 474, 2757

\bibitem[{{Heger} {et~al.}(2003){Heger}, {Fryer}, {Woosley}, {Langer}, \& {Hartmann}}]{Heger2003}
{Heger}, A., {Fryer}, C.~L., {Woosley}, S.~E., {Langer}, N., \& {Hartmann}, D.~H. 2003, \apj, 591, 288

\bibitem[{{Heger} \& {Woosley}(2002)}]{Heger2002}
{Heger}, A. \& {Woosley}, S.~E. 2002, \apj, 567, 532

\bibitem[{{Hernquist} \& {Katz}(1989)}]{Hernquist_Katz1989}
{Hernquist}, L. \& {Katz}, N. 1989, \apjs, 70, 419

\bibitem[{{Hosokawa} {et~al.}(2016){Hosokawa}, {Hirano}, {Kuiper}, {Yorke}, {Omukai}, \& {Yoshida}}]{hosokawa2016}
{Hosokawa}, T., {Hirano}, S., {Kuiper}, R., {et~al.} 2016, \apj, 824, 119

\bibitem[{{Hosokawa} \& {Omukai}(2009)}]{Hosokawa09}
{Hosokawa}, T. \& {Omukai}, K. 2009, \apj, 703, 1810

\bibitem[{{Hosokawa} {et~al.}(2012){Hosokawa}, {Omukai}, \& {Yorke}}]{Hosokawa12}
{Hosokawa}, T., {Omukai}, K., \& {Yorke}, H.~W. 2012, \apj, 756, 93

\bibitem[{{Hosokawa} {et~al.}(2013){Hosokawa}, {Yorke}, {Inayoshi}, {Omukai}, \& {Yoshida}}]{Hosokawa2013}
{Hosokawa}, T., {Yorke}, H.~W., {Inayoshi}, K., {Omukai}, K., \& {Yoshida}, N. 2013, \apj, 778, 178

\bibitem[{{Hubber} {et~al.}(2013){Hubber}, {Walch}, \& {Whitworth}}]{Hubber13}
{Hubber}, D.~A., {Walch}, S., \& {Whitworth}, A.~P. 2013, \mnras, 430, 3261

\bibitem[{{Inayoshi} {et~al.}(2018){Inayoshi}, {Ostriker}, {Haiman}, \& {Kuiper}}]{2018inayoshi}
{Inayoshi}, K., {Ostriker}, J.~P., {Haiman}, Z., \& {Kuiper}, R. 2018, \mnras, 476, 1412

\bibitem[{{Johnson} \& {Bromm}(2007)}]{Johnson2007}
{Johnson}, J.~L. \& {Bromm}, V. 2007, \mnras, 374, 1557

\bibitem[{{Katz} {et~al.}(2015){Katz}, {Sijacki}, \& {Haehnelt}}]{Katz2015}
{Katz}, H., {Sijacki}, D., \& {Haehnelt}, M.~G. 2015, \mnras, 451, 2352

\bibitem[{{Klessen}(2019)}]{Klessen19}
{Klessen}, R. 2019, {Formation of the first stars}, ed. M.~{Latif} \& D.~{Schleicher}, 67--97

\bibitem[{{Koushiappas} {et~al.}(2004){Koushiappas}, {Bullock}, \& {Dekel}}]{Koushiappas2004}
{Koushiappas}, S.~M., {Bullock}, J.~S., \& {Dekel}, A. 2004, \mnras, 354, 292

\bibitem[{{Latif} \& {Schleicher}(2015)}]{LatifSchleicher2015}
{Latif}, M.~A. \& {Schleicher}, D.~R.~G. 2015, \aap, 578, A118

\bibitem[{{Latif} {et~al.}(2013){Latif}, {Schleicher}, {Schmidt}, \& {Niemeyer}}]{Latif2013}
{Latif}, M.~A., {Schleicher}, D.~R.~G., {Schmidt}, W., \& {Niemeyer}, J. 2013, \mnras, 433, 1607

\bibitem[{{Leigh} {et~al.}(2017){Leigh}, {Geller}, {Shara}, {Garland}, {Clees-Baron}, \& {Ahmed}}]{Leigh2017}
{Leigh}, N. W.~C., {Geller}, A.~M., {Shara}, M.~M., {et~al.} 2017, \mnras, 471, 1830

\bibitem[{{Lombardi} {et~al.}(2002){Lombardi}, {Warren}, {Rasio}, {Sills}, \& {Warren}}]{2002lombardi}
{Lombardi}, James~C., J., {Warren}, J.~S., {Rasio}, F.~A., {Sills}, A., \& {Warren}, A.~R. 2002, \apj, 568, 939

\bibitem[{{Lupi} {et~al.}(2014){Lupi}, {Colpi}, {Devecchi}, {Galanti}, \& {Volonteri}}]{Lupi14}
{Lupi}, A., {Colpi}, M., {Devecchi}, B., {Galanti}, G., \& {Volonteri}, M. 2014, \mnras, 442, 3616

\bibitem[{{Maud} {et~al.}(2019){Maud}, {Cesaroni}, {Kumar}, {Rivilla}, {Ginsburg}, {Klaassen}, {Harsono}, {S{\'a}nchez-Monge}, {Ahmadi}, {Allen}, {Beltr{\'a}n}, {Beuther}, {Galv{\'a}n-Madrid}, {Goddi}, {Hoare}, {Hogerheijde}, {Johnston}, {Kuiper}, {Moscadelli}, {Peters}, {Testi}, {van der Tak}, \& {de Wit}}]{2019Maud_observation}
{Maud}, L.~T., {Cesaroni}, R., {Kumar}, M.~S.~N., {et~al.} 2019, \aap, 627, L6

\bibitem[{{McMillan} \& {Hut}(1996)}]{McMillan96}
{McMillan}, S. L.~W. \& {Hut}, P. 1996, \apj, 467, 348

\bibitem[{{Mortlock} {et~al.}(2011){Mortlock}, {Warren}, {Venemans}, {Patel}, {Hewett}, {McMahon}, {Simpson}, {Theuns}, {Gonz{\'a}les-Solares}, {Adamson}, {Dye}, {Hambly}, {Hirst}, {Irwin}, {Kuiper}, {Lawrence}, \& {R{\"o}ttgering}}]{Mortlock2011}
{Mortlock}, D.~J., {Warren}, S.~J., {Venemans}, B.~P., {et~al.} 2011, \nat, 474, 616

\bibitem[{{Oliva} \& {Kuiper}(2020)}]{2020discOliva}
{Oliva}, G.~A. \& {Kuiper}, R. 2020, \aap, 644, A41

\bibitem[{{Omukai}(2001)}]{Omukai2001}
{Omukai}, K. 2001, \apj, 546, 635

\bibitem[{{Omukai} {et~al.}(2008){Omukai}, {Schneider}, \& {Haiman}}]{Omukai2008}
{Omukai}, K., {Schneider}, R., \& {Haiman}, Z. 2008, \apj, 686, 801

\bibitem[{{Omukai} {et~al.}(2005){Omukai}, {Tsuribe}, {Schneider}, \& {Ferrara}}]{Omukai2005}
{Omukai}, K., {Tsuribe}, T., {Schneider}, R., \& {Ferrara}, A. 2005, \apj, 626, 627

\bibitem[{{Onoue} {et~al.}(2019){Onoue}, {Kashikawa}, {Matsuoka}, {Kato}, {Izumi}, {Nagao}, {Strauss}, {Harikane}, {Imanishi}, {Ito}, {Iwasawa}, {Kawaguchi}, {Lee}, {Noboriguchi}, {Suh}, {Tanaka}, \& {Toba}}]{Onoue2019}
{Onoue}, M., {Kashikawa}, N., {Matsuoka}, Y., {et~al.} 2019, \apj, 880, 77

\bibitem[{{Pelupessy} {et~al.}(2004){Pelupessy}, {van der Werf}, \& {Icke}}]{Pelupessy2004}
{Pelupessy}, F.~I., {van der Werf}, P.~P., \& {Icke}, V. 2004, \aap, 422, 55

\bibitem[{{Pelupessy} {et~al.}(2013){Pelupessy}, {van Elteren}, {de Vries}, {McMillan}, {Drost}, \& {Portegies Zwart}}]{AMUSE_Pelupessy13}
{Pelupessy}, F.~I., {van Elteren}, A., {de Vries}, N., {et~al.} 2013, \aap, 557, A84

\bibitem[{{Peters} {et~al.}(2010){Peters}, {Klessen}, {Mac Low}, \& {Banerjee}}]{2010peters}
{Peters}, T., {Klessen}, R.~S., {Mac Low}, M.-M., \& {Banerjee}, R. 2010, \apj, 725, 134

\bibitem[{{Plummer}(1911)}]{plummer1911}
{Plummer}, H.~C. 1911, \mnras, 71, 460

\bibitem[{{Portegies Zwart} \& {McMillan}(2018)}]{Portegies2018}
{Portegies Zwart}, S. \& {McMillan}, S. 2018, {Astrophysical Recipes; The art of AMUSE}

\bibitem[{{Portegies Zwart} {et~al.}(2009){Portegies Zwart}, {McMillan}, {Harfst}, {Groen}, {Fujii}, {Nuall{\'a}in}, {Glebbeek}, {Heggie}, {Lombardi}, {Hut}, {Angelou}, {Banerjee}, {Belkus}, {Fragos}, {Fregeau}, {Gaburov}, {Izzard}, {Juri{\'c}}, {Justham}, {Sottoriva}, {Teuben}, {van Bever}, {Yaron}, \& {Zemp}}]{AMUSE_Portegies09}
{Portegies Zwart}, S., {McMillan}, S., {Harfst}, S., {et~al.} 2009, \na, 14, 369

\bibitem[{{Portegies Zwart} {et~al.}(2013){Portegies Zwart}, {McMillan}, {van Elteren}, {Pelupessy}, \& {de Vries}}]{AMUSE_Portegies13}
{Portegies Zwart}, S., {McMillan}, S.~L.~W., {van Elteren}, E., {Pelupessy}, I., \& {de Vries}, N. 2013, Computer Physics Communications, 184, 456

\bibitem[{{Prole} {et~al.}(2024){Prole}, {Regan}, {Glover}, {Klessen}, {Priestley}, \& {Clark}}]{2024prole}
{Prole}, L.~R., {Regan}, J.~A., {Glover}, S. C.~O., {et~al.} 2024, \aap, 685, A31

\bibitem[{{Reed} {et~al.}(2019){Reed}, {Banerji}, {Becker}, {Hewett}, {Martini}, {McMahon}, {Pons}, {Rauch}, {Abbott}, {Allam}, {Annis}, {Avila}, {Bertin}, {Brooks}, {Buckley-Geer}, {Carnero Rosell}, {Carrasco Kind}, {Carretero}, {Castander}, {Cunha}, {D'Andrea}, {da Costa}, {De Vicente}, {Desai}, {Diehl}, {Doel}, {Evrard}, {Flaugher}, {Frieman}, {Garc{\'\i}a-Bellido}, {Gaztanaga}, {Gruen}, {Gschwend}, {Gutierrez}, {Hollowood}, {Honscheid}, {Hoyle}, {James}, {Kuehn}, {Lahav}, {Lima}, {Maia}, {Marshall}, {Miquel}, {Ogand o}, {Plazas}, {Roodman}, {Sanchez}, {Scarpine}, {Schubnell}, {Serrano}, {Sevilla-Noarbe}, {Smith}, {Smith}, {Sobreira}, {Suchyta}, {Swanson}, {Tarle}, {Thomas}, {Tucker}, \& {Vikram}}]{Reed2019}
{Reed}, S.~L., {Banerji}, M., {Becker}, G.~D., {et~al.} 2019, \mnras, 487, 1874

\bibitem[{{Rees}(1984)}]{Rees1984}
{Rees}, M.~J. 1984, \araa, 22, 471

\bibitem[{{Regan}(2023)}]{2023regan}
{Regan}, J. 2023, The Open Journal of Astrophysics, 6, 12

\bibitem[{{Regan} \& {Downes}(2018{\natexlab{a}})}]{2018regan2}
{Regan}, J.~A. \& {Downes}, T.~P. 2018{\natexlab{a}}, \mnras, 475, 4636

\bibitem[{{Regan} \& {Downes}(2018{\natexlab{b}})}]{2018regan}
{Regan}, J.~A. \& {Downes}, T.~P. 2018{\natexlab{b}}, \mnras, 478, 5037

\bibitem[{{Reinoso} {et~al.}(2023){Reinoso}, {Klessen}, {Schleicher}, {Glover}, \& {Solar}}]{Reinoso2023}
{Reinoso}, B., {Klessen}, R.~S., {Schleicher}, D., {Glover}, S. C.~O., \& {Solar}, P. 2023, \mnras, 521, 3553

\bibitem[{{Reinoso} {et~al.}(2018){Reinoso}, {Schleicher}, {Fellhauer}, {Klessen}, \& {Boekholt}}]{Reinoso18}
{Reinoso}, B., {Schleicher}, D.~R.~G., {Fellhauer}, M., {Klessen}, R.~S., \& {Boekholt}, T.~C.~N. 2018, \aap, 614, A14

\bibitem[{{Reinoso} {et~al.}(2020){Reinoso}, {Schleicher}, {Fellhauer}, {Leigh}, \& {Klessen}}]{Reinoso2020}
{Reinoso}, B., {Schleicher}, D.~R.~G., {Fellhauer}, M., {Leigh}, N.~W.~C., \& {Klessen}, R.~S. 2020, \aap, 639, A92

\bibitem[{{Riaz} {et~al.}(2018){Riaz}, {Bovino}, {Vanaverbeke}, \& {Schleicher}}]{Rafeel18}
{Riaz}, R., {Bovino}, S., {Vanaverbeke}, S., \& {Schleicher}, D.~R.~G. 2018, \mnras, 479, 667

\bibitem[{{Sakurai} {et~al.}(2016){Sakurai}, {Vorobyov}, {Hosokawa}, {Yoshida}, {Omukai}, \& {Yorke}}]{2016sakurai}
{Sakurai}, Y., {Vorobyov}, E.~I., {Hosokawa}, T., {et~al.} 2016, \mnras, 459, 1137

\bibitem[{{Sakurai} {et~al.}(2019){Sakurai}, {Yoshida}, \& {Fujii}}]{Sakurai2019}
{Sakurai}, Y., {Yoshida}, N., \& {Fujii}, M.~S. 2019, \mnras, 484, 4665

\bibitem[{{Sakurai} {et~al.}(2017){Sakurai}, {Yoshida}, {Fujii}, \& {Hirano}}]{Sakurai2017}
{Sakurai}, Y., {Yoshida}, N., {Fujii}, M.~S., \& {Hirano}, S. 2017, \mnras, 472, 1677

\bibitem[{{Sassano} {et~al.}(2021){Sassano}, {Schneider}, {Valiante}, {Inayoshi}, {Chon}, {Omukai}, {Mayer}, \& {Capelo}}]{Sassano2021}
{Sassano}, F., {Schneider}, R., {Valiante}, R., {et~al.} 2021, \mnras, 506, 613

\bibitem[{{Schleicher} {et~al.}(2013){Schleicher}, {Palla}, {Ferrara}, {Galli}, \& {Latif}}]{Schleicher2013}
{Schleicher}, D. R.~G., {Palla}, F., {Ferrara}, A., {Galli}, D., \& {Latif}, M. 2013, \aap, 558, A59

\bibitem[{{Schleicher} {et~al.}(2023){Schleicher}, {Reinoso}, \& {Klessen}}]{Dominik_super2023}
{Schleicher}, D. R.~G., {Reinoso}, B., \& {Klessen}, R.~S. 2023, \mnras, 521, 3972

\bibitem[{{Schleicher} {et~al.}(2022){Schleicher}, {Reinoso}, {Latif}, {Klessen}, {Vergara}, {Das}, {Alister}, {D{\'\i}az}, \& {Solar}}]{Schleicher2022}
{Schleicher}, D.~R.~G., {Reinoso}, B., {Latif}, M., {et~al.} 2022, \mnras, 512, 6192

\bibitem[{{Schleicher} {et~al.}(2010){Schleicher}, {Spaans}, \& {Glover}}]{Schleicher2010}
{Schleicher}, D. R.~G., {Spaans}, M., \& {Glover}, S. C.~O. 2010, \apjl, 712, L69

\bibitem[{{Schneider} {et~al.}(2003){Schneider}, {Ferrara}, {Salvaterra}, {Omukai}, \& {Bromm}}]{Schneider2003}
{Schneider}, R., {Ferrara}, A., {Salvaterra}, R., {Omukai}, K., \& {Bromm}, V. 2003, \nat, 422, 869

\bibitem[{{Schneider} {et~al.}(2006){Schneider}, {Omukai}, {Inoue}, \& {Ferrara}}]{Schneider2006}
{Schneider}, R., {Omukai}, K., {Inoue}, A.~K., \& {Ferrara}, A. 2006, \mnras, 369, 1437

\bibitem[{{Sharda} {et~al.}(2020){Sharda}, {Federrath}, \& {Krumholz}}]{Sharda2020}
{Sharda}, P., {Federrath}, C., \& {Krumholz}, M.~R. 2020, \mnras, 497, 336

\bibitem[{{Sills} \& {Bailyn}(1999)}]{1999sills}
{Sills}, A. \& {Bailyn}, C.~D. 1999, \apj, 513, 428

\bibitem[{{Sills} {et~al.}(2000){Sills}, {Bailyn}, {Edmonds}, \& {Gilliland}}]{2000sills}
{Sills}, A., {Bailyn}, C.~D., {Edmonds}, P.~D., \& {Gilliland}, R.~L. 2000, \apj, 535, 298

\bibitem[{{Smith} {et~al.}(2018){Smith}, {Regan}, {Downes}, {Norman}, {O'Shea}, \& {Wise}}]{Smith2018}
{Smith}, B.~D., {Regan}, J.~A., {Downes}, T.~P., {et~al.} 2018, \mnras, 480, 3762

\bibitem[{{Solar} {et~al.}(2022){Solar}, {Schleicher}, {Reinoso}, \& {Klessen}}]{paulo2022}
{Solar}, P.~A., {Schleicher}, D.~R.~G., {Reinoso}, B., \& {Klessen}, R.~S. 2022, Boletin de la Asociacion Argentina de Astronomia La Plata Argentina, 63, 277

\bibitem[{{Stacy} {et~al.}(2016){Stacy}, {Bromm}, \& {Lee}}]{Stacy2016}
{Stacy}, A., {Bromm}, V., \& {Lee}, A.~T. 2016, \mnras, 462, 1307

\bibitem[{{Suazo} {et~al.}(2019){Suazo}, {Prieto}, {Escala}, \& {Schleicher}}]{Suazo2019}
{Suazo}, M., {Prieto}, J., {Escala}, A., \& {Schleicher}, D. R.~G. 2019, \apj, 885, 127

\bibitem[{{Susa} {et~al.}(2014){Susa}, {Hasegawa}, \& {Tominaga}}]{Susa2014}
{Susa}, H., {Hasegawa}, K., \& {Tominaga}, N. 2014, \apj, 792, 32

\bibitem[{{Tagawa} {et~al.}(2020){Tagawa}, {Haiman}, \& {Kocsis}}]{Tagawa2020}
{Tagawa}, H., {Haiman}, Z., \& {Kocsis}, B. 2020, \apj, 892, 36

\bibitem[{{Toomre}(1964)}]{toomre}
{Toomre}, A. 1964, \apj, 139, 1217

\bibitem[{{Trinca} {et~al.}(2022){Trinca}, {Schneider}, {Valiante}, {Graziani}, {Zappacosta}, \& {Shankar}}]{Trinca2022}
{Trinca}, A., {Schneider}, R., {Valiante}, R., {et~al.} 2022, \mnras, 511, 616

\bibitem[{{Vergara} {et~al.}(2022){Vergara}, {Escala}, {Schleicher}, \& {Reinoso}}]{Vergara2022}
{Vergara}, M.~C., {Escala}, A., {Schleicher}, D.~R.~G., \& {Reinoso}, B. 2022, arXiv e-prints, arXiv:2209.15066

\bibitem[{{Vergara} {et~al.}(2023){Vergara}, {Escala}, {Schleicher}, \& {Reinoso}}]{2023vergara}
{Vergara}, M.~C., {Escala}, A., {Schleicher}, D.~R.~G., \& {Reinoso}, B. 2023, \mnras, 522, 4224

\bibitem[{{Vergara} {et~al.}(2021){Vergara}, {Schleicher}, {Boekholt}, {Reinoso}, {Fellhauer}, {Klessen}, \& {Leigh}}]{Vergara21}
{Vergara}, M.~Z.~C., {Schleicher}, D.~R.~G., {Boekholt}, T.~C.~N., {et~al.} 2021, \aap, 649, A160

\bibitem[{{Volonteri}(2010)}]{Volonteri10}
{Volonteri}, M. 2010, \aapr, 18, 279

\bibitem[{{Wang} {et~al.}(2021){Wang}, {Yang}, {Fan}, {Hennawi}, {Barth}, {Banados}, {Bian}, {Boutsia}, {Connor}, {Davies}, {Decarli}, {Eilers}, {Farina}, {Green}, {Jiang}, {Li}, {Mazzucchelli}, {Nanni}, {Schindler}, {Venemans}, {Walter}, {Wu}, \& {Yue}}]{Wang21}
{Wang}, F., {Yang}, J., {Fan}, X., {et~al.} 2021, \apjl, 907, L1

\bibitem[{{Wise} {et~al.}(2008){Wise}, {Turk}, \& {Abel}}]{Wise2008}
{Wise}, J.~H., {Turk}, M.~J., \& {Abel}, T. 2008, \apj, 682, 745

\bibitem[{Woods {et~al.}(2019)Woods, Agarwal, Bromm, Bunker, Chen, Chon, Ferrara, Glover, Haemmerlé, Haiman, \& et~al.}]{review_woods19}
Woods, T.~E., Agarwal, B., Bromm, V., {et~al.} 2019, Publications of the Astronomical Society of Australia, 36, e027

\bibitem[{{Wu} {et~al.}(2015){Wu}, {Wang}, {Fan}, {Yi}, {Zuo}, {Bian}, {Jiang}, {McGreer}, {Wang}, {Yang}, {Yang}, {Thompson}, \& {Beletsky}}]{Wu15Nature}
{Wu}, X.-B., {Wang}, F., {Fan}, X., {et~al.} 2015, \nat, 518, 512

\end{thebibliography}

\begin{appendix}
\onecolumn
\section{Edge-on view}\label{appendix_edge_on}
In this appendix, we present the density projection of the edge-on view, similar to Fig. \ref{fig:morphology_plot}.

\begin{figure*}[h!]
    \centering
	\includegraphics[width=17cm]{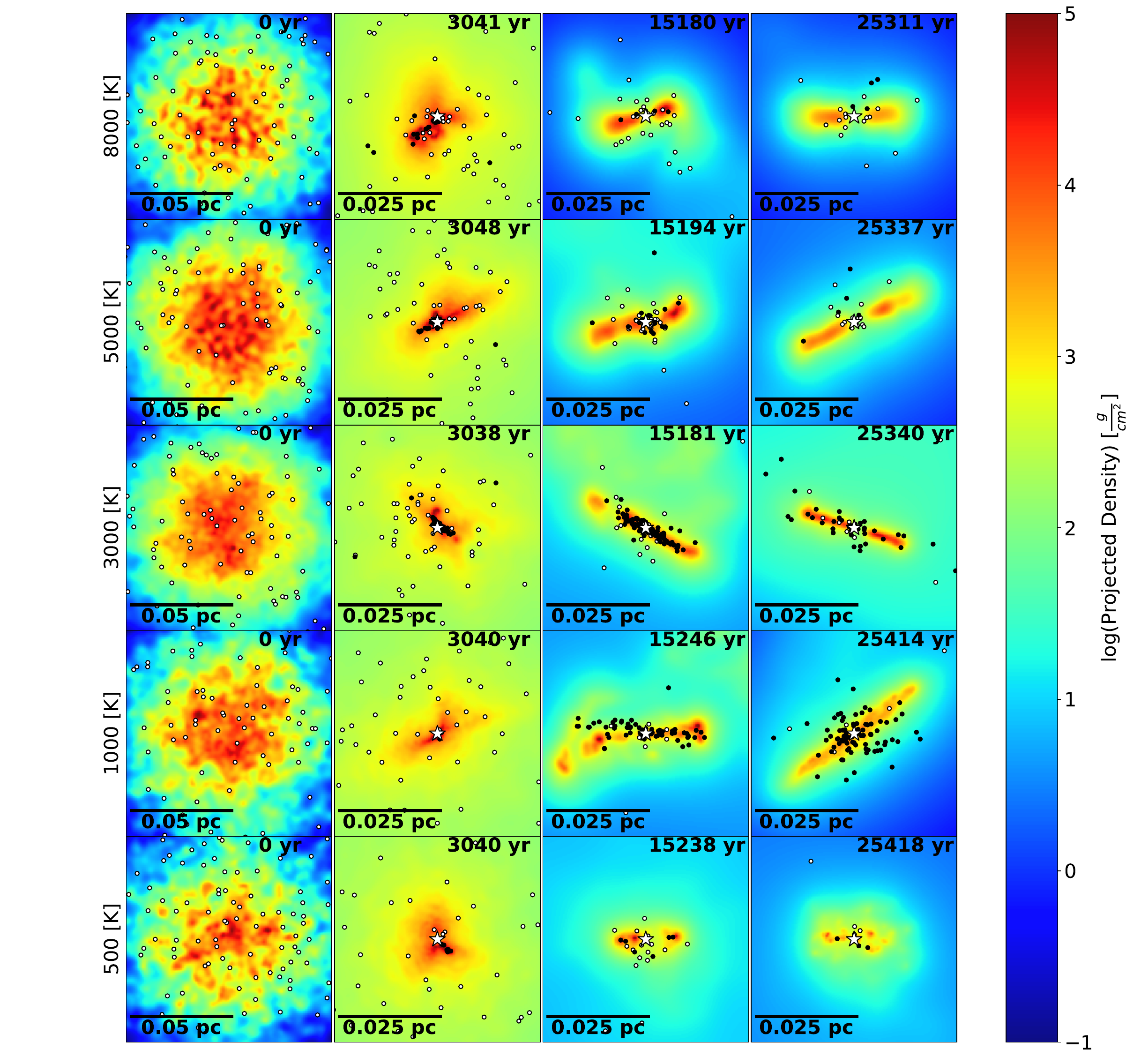}
    \caption{Density projections of the edge-on view of a primordial gas cloud with an embedded protostar cluster for simulations with an initial temperature of $500~\mathrm{K}-8000~\mathrm{K}$ at different times. The white dots are the protostars, black dots are the formed stars and the white star is the MMO. The clusters evolve in time from left to right and from the warmer to colder simulations from the top to the bottom panels.
    }
    \label{fig:morphology_plot_edge}
\end{figure*}

\end{appendix}

\end{document}